# Narwhal 上に構築された Bullshark ラウンドベース DAG コンセンサスプロトコルの理論と実用におけるワークフロー分析

田中　優成[†,††]

† Monas
†† 関西大学　〒564-8680 大阪府吹田市山手町 3 丁目 3 番 35 号
E-mail: †yusurf.osaru@gmail.com

**あらまし**　ビザンチン障害耐性を持つコンセンサスを最適化したラウンドベース DAG は極めて高いパフォーマンスを誇るが，その歴史は浅く，技術的優位性の理解と普及が進んでいない．また，コンセンサスプロトコルの研究は学術界産業界ともに活発であるが，実行アルゴリズムを考慮していないため，実際のパフォーマンスは不明瞭であり，特に理論的なプロトコルは実用時の性能は評価できない．Bullshark は，Narwhal メンプールプロトコル上に構築された BFT コンセンサスプロトコルであり，ラウンドベース DAG を採用している．この組み合わせにより，297,000 tx/s のスループットと約 2 秒の低レイテンシという現時点で最高水準のパフォーマンスを達成している．本研究では，Bullshark におけるユーザによるトランザクションの送信からブロックチェーンコミットまでのアルゴリズムを各レイヤーごとに関数レベルで分析した．この分析を通じて，Narwhal と Bullshark の技術的特徴と相互関係を明らかにし，そのメカニズムの理論と実用の両面を紐解いた．今後は，ビザンチン環境下でのパフォーマンス改善や CAP 定理におけるトレードオフの最適化が期待される．

# Bullshark on Narwhal: Implementation-level Workflow Analysis of Round-based DAG Consensus in Theory and Practice.

Yusei TANAKA[†,††]

† Monas
†† Kansai University　3-3-35 Yamate-cho, Suita-shi, Osaka, 564-8680 Japan
E-mail: †yusurf.osaru@gmail.com

**Abstract**　Round-based DAGs enable high-performance Byzantine fault-tolerant consensus, yet their technical advantages remain underutilized due to their short history. While research on consensus protocols is active in both academia and industry, many studies overlook implementation-level algorithms, leaving actual performance unclear—particularly for theoretical protocols whose practical performance cannot often be evaluated. Bullshark, a Round-based DAG BFT protocol on Narwhal mempool, achieves optimal performance: 297,000 transactions per second with 2-second latency. We analyze the algorithm's workflow, from transaction submission to blockchain commitment, breaking it down layer by layer at the functional level and delineating the key features and interactions of the Bullshark and Narwhal components. Future work aims to improve performance in Byzantine fault environments and optimize trade-offs in the CAP theorem.

## 1. 序　　論

Bullshark とは，Narwhal メンプール上に構築された，部分同期に最適化した DAG ベースの BFT コンセンサスプロトコルである．パブリックパーミッションレスブロックチェーンにおいて，障害が起きてもシステムを動作させ続けるために，過去 10 年間，ビザンチンフォールトトレランス性（BFT）が注目されてきた．その中で，ブロックチェーンのスケラビリティ問題 [1] で蔑ろにされてきたスループットを拡大するために BFT コンセンサスプロトコルに Directed Acyclic Graph（DAG）トポロジー

を導入したコンセンサスが提案され続けている．（図 1）これは，通信や計算，ストレージのオーバーヘッドを少なくすることで，パフォーマンスを向上させる方法で，より多くのトランザクションを処理することを可能にしたものである [2]．概念としては，2015 年，S.D.Lerner によって，DAG-chain [3] として初めて提唱された．ブロックレベルの DAG としては，GHOST [4] で初めて提案された．

Bullshark は実際のプロダクション環境下で使用され，実測値としてそのパフォーマンスが 297000+TPS でブロックチェーン産業史上最大のスループットを記録している．これは，パフォーマンスが高いことで知られる Solana でも理論的に最大 65,000 TPS [5] であることを考えると，ブロックチェーンのパフォーマンスの大部分はコンセンサスプロトコルの性能で決まるため，Bullshark を理解することは現在のブロックチェーンネットワークのパフォーマンスにおいて最大可能性について理解することにつながる．また，Bullshark と Narwhal はオープンソースで実装されており，理論的な分析にとどまることはなく，プロダクション環境を直接分析することができる．

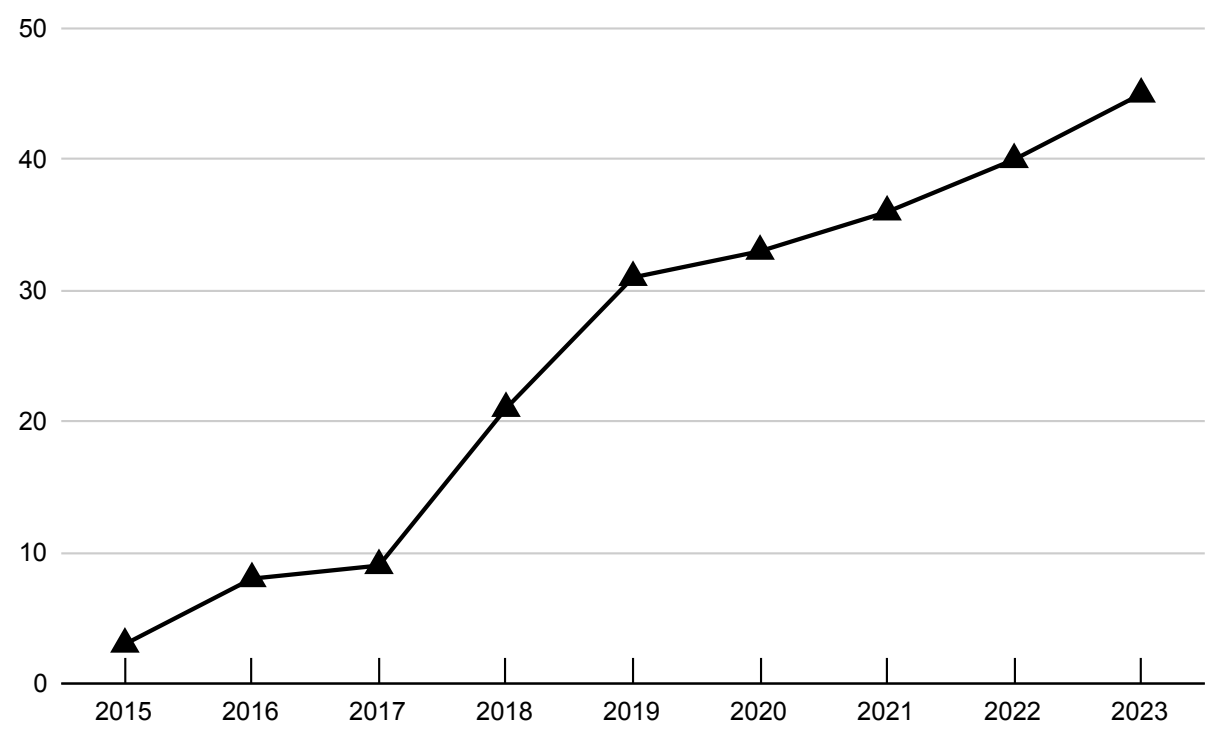

図 1: これまでに提案された DAG ベースのコンセンサスプロトコルの総数

DAG ベースのコンセンサスプロトコルの分析をした関連研究 [2],[6] では，理論的な分類や比較にとどまり，実装レベルでの詳細な分析をしておらず，実際のブロックチェーンネットワークでの動作特性が検証されていない．[2] では，既存および進行中の研究をレビューして分析して概要を伝えることが目的となっており，各コンセンサスプロトコルの詳細なアルゴリズムについては説明されていない．[7] においては，抽象的なモデル設計に焦点されており，実際の具体的な実装方法や実用面での可能性についての議論は不十分である．これらの研究は理論やモデル化といったコンセンサスプロトコルの設計については言及にとどまっており，実際の実行アルゴリズムについては全く言及されていない．コンセンサスのパフォーマンスが良いことが理論上示されたとしてもトランザクションの実行によってパフォーマンスが悪くなることがあり，それは無視できない影響を与える [8] ため，理論やモデルの検証では実用面では不十分であり，実際のネットワークのパフォーマンスを理解できない．

本研究はコンセンサスプロトコルの順序付けのみに焦点を当てるのではなく，今までなされなかったトランザクションの送信処理から実行アルゴリズムまでを実装レベルの詳細な分析に焦点を置き，理論と実践のギャップを埋めている．Bullshark と Narwhal 上におけるトランザクション送信からブロックチェーンコミットまでのエンドツーエンドのトランザクション処理フローの全プロセスを示し，その中での技術的特徴と相互関係，297,000 TPS という高スループットを達成するメカニズムを明らかなものとしている．さらに，コンセンサスプロトコルの分野で DAG を活用して BAB 問題や理論的課題，DAG の技術的課題などの主要な課題をどのように解決するのか，それはどのような技術的特徴によって達成されているのかを示している．私たちはブロックチェーンのパフォーマンスの鍵を握るコンセンサスプロトコルの中で市場で最も高いパフォーマンスを実測した Bullshark と Narwhal を関数レベルでネットワークレイヤー，コンセンサスレイヤー，実行レイヤーでのアルゴリズムを分析した結果，今後のより高いパフォーマンスのコンセンサスを実装するための重要な歴史と背景も示している．

2. で，コンセンサスプロトコルの歴史を示し，どのような技術的背景によって DAG が注目されたのかを示す．3. で，その DAG の技術的特徴と特性を明確する．4. と 5. で Narwhal と Bullshark の概要と解決している課題や特性を特定し，Narwhal ではノードのアーキテクチャについて，Bullshark ではパフォーマンスに寄与する重要な技術的特徴について，理論的なコンセンサスアルゴリズムについて理解する．6. で，Narwhal と Bullshark でのトランザクション送信からブロックチェーンコミットまでのネットワークレイヤーからコンセンサスレイヤー，実行レイヤーにまたがる詳細なアルゴリズムをプロダクション環境での関数レベルで分析し，そのメカニズムを理解する．

## 2. コンセンサスプロトコルの歴史

コンセンサスプロトコルとは，ブロックチェーンネットワーク上でトランザクションと呼ばれるプログラムやタイムスタンプ，署名などの取引情報の真正性（Authenticity）をノード間で検証し，ノード全体で 1 つの状態の共有台帳に合意形成するシステムまたはルールのことである．また，パブリックパーミッションレスブロックチェーンでは，取引における真正性は公開されており，誰もが検証することができることにより，監査可能性（Auditability）が保証されている．一般的に分散ネットワーク上ではしばしば障害が起きる．この時，クラッシュやタイムアウトなど時間内に反応することができない状態となることもあるが，ブロックチェーンは障害時においても動作する堅牢なシステムでなければならない．また，コンセンサスプロトコルはこの性質を保証することが求められる．

**safety.** 安全性とは，コンセンサスプロトコルにおいて，悪意のあるまたは遅延している不正なノードが存在していたとしても，正直なノードによって常に一貫した状態に合意形成される性質のことである．これはデータの信頼性や可用性，一貫性を満たすための前提条件となる．

**Liveness.** 生存性とは，コンセンサスプロトコルにおいて，正直なノードは正しく動作している限り，コミットまで進行を続

け，合意形成を達成することができるという性質のことである．

1980 年，ビザンチン将軍問題（The Byzantines General’s Problem）のコンセンサス問題として定式化された [9]．ビザンチン障害耐性（Byzantine Fault Tolerance, BFT）とは，システムの一部に問題が生じても全体が機能停止するということなく，システムを維持し続ける耐性のことである．そして，1982 年，3 分の 2 以上が honest に正しい行動をするとき，合意システムは維持可能であることが示された [10]．しかしながら，BFT レプリケーションプロトコルは，コストがかかり実用的ではないとされていた．1999 年，B.Liskov らによって，初めて実用可能な BFT コンセンサスアルゴリズムである PBFT [11] が提案され，高速処理が可能となった．しかしノードが増えると投票に対する通信量が増加し，スケーリングに問題を抱えていた．2008 年にはビットコイン [12] がこれを応用し，Proof of Work（PoW）メカニズム [12] を設計した．PoW を基盤としたナカモト・コンセンサス [12] は信頼性や公正性などを満たす各ノードによる分散的な合意形成を可能とし，ブロックチェーン産業を大きく拡大させる要因の一つとなった．サトシナカモトによれば，PoW チェーンはビザンチン将軍問題の解決策の一つなのである [13]．その後，PBFT [11] は SBFT [14] や Mir-BFT [15], Cosmos [16] の Tendermint [17] など多くのバリアント [18] [19] が提案された．これらは非常に高いパフォーマンスを発揮するものの，攻撃やノード障害による性能劣化が顕著であった [20], 2018 年に，VMWare Research は，PBFT [11] を大幅に改良した，HotStuff（HS）[21] を開発し，Tendermint [17] とあらゆるネットワーク条件で同じ計算複雑性を保ちながら，3 チェーン化することで，堅牢性と安全性，信頼性を高め，ピーク性能で 70,000 tx/s を達成することが確認された [22]．現在最も広く知られる Ethereum の The Merge [23] 以前では，25tx/s 程度 [24] のスループットしか達成できていないことを考えると非常に高いパフォーマンスであると言える．Facebook 社はパーミッションドブロックチェーンであるものの，HS を LibraBFT [25] として採用し，Diem へのリブランディング以降は DiemBFT [26] とした．

ブロックチェーンのスケラビリティ問題 [1] を解決するためには多くのスケーリング技術が提案されている．スケーリング技術で一般的に利用されているものとして，レイヤー 2 プロトコル [27] やサイドチェーン [28] などが挙げられる．このうち DAG（Directed Acyclic Graph）を活用したものがある．これは，通信や計算，ストレージのオーバーヘッドを少なくすることで，パフォーマンスを向上させる方法で，より多くのトランザクションを処理することを可能にした [2]．概念としては，2015 年，S.D.Lerner によって，DAG-chain [3] として初めて提唱された．ブロックレベルの DAG としては，GHOST [4] で初めて提案された．

## 3. Directed Acyclic Graph（DAG）

### 3.1 DAG とは

DAG とは，Directed Acyclic Graph の略で，日本語で有向非循環グラフを意味する．アルゴリズムとデータ構造の分野のグラフ理論の中の一般的に利用されるプリミティブなデータ構造である．2 つの頂点（vertex）と辺（edge）で構成され，この辺は常にある頂点から別の頂点への方向性を持ち，サイクルを持たない．あなたがエンジニアであるならば GitHub でのバージョン管理が DAG によって構築されていると言われれば，イメージしやすいだろう．一般的には，家系図のように，どの先祖から出発しても自分自身になるといったことをイメージされたい．図 2 の例に従えば，a の祖父からスタートしても必ず時系列的には e の自分に到達するということである．

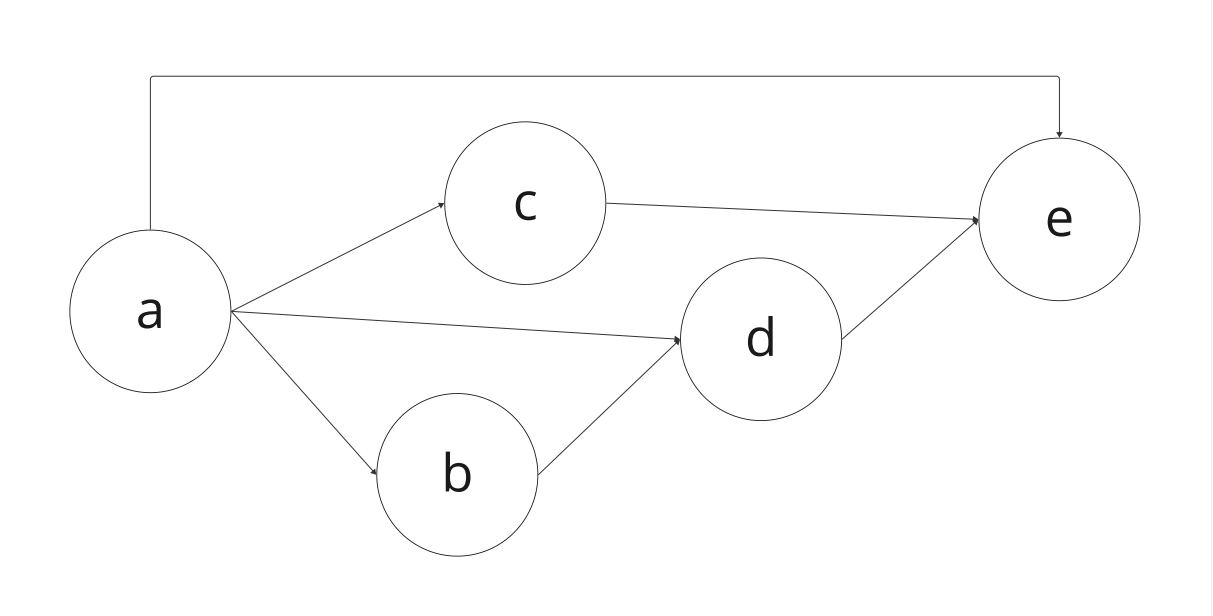

図 2: DAG の例

### 3.2 DAG と伝統的ブロックチェーンとの比較 [29]

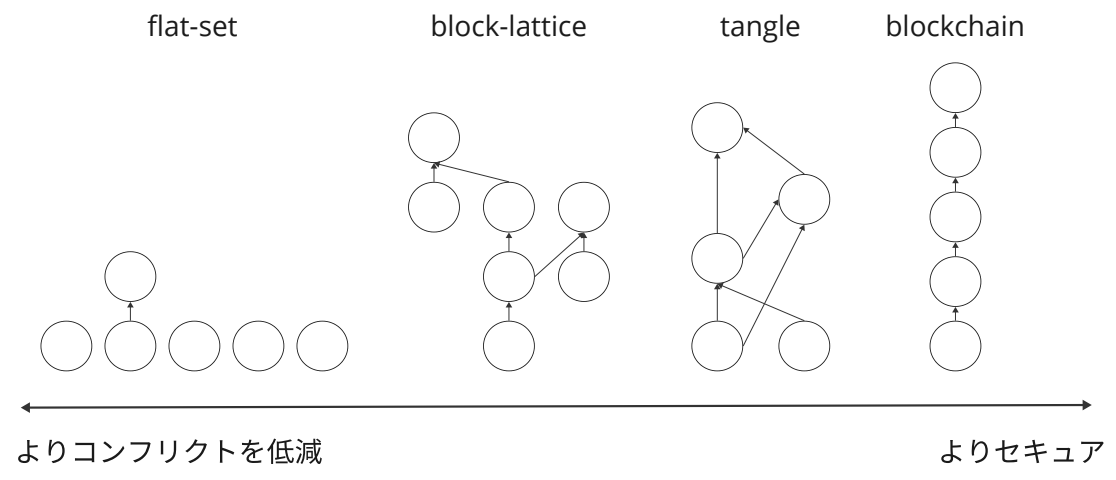

図 3: DAG と伝統的ブロックチェーンとの比較図

この図は，DAG 構造を一次元のリスト構造と要素間の関係や階層を持たない集合とを比較したものである．セキュリティが左から右に徐々に改善されているが，同時にパフォーマンスが徐々に低下しているため、TPS が小さくなる．左端の flat-set はプライベートブロックチェーンなど集中型システムで一般的に使用される基本構造であり，右端の blockchain は典型的なブロックチェーン構造であり，その間がどちらも DAG 構造である．blockchain 構造は，Ethereum 等でブロックを線形に構築する．これはメンプールにおいてもいうことができ，多くのブロックチェーンは配列データ構造で，単純にトランザクションを線形に追加されるだけである．DAG 構造である Block-Lattice と Tangle は，右端の blockchain と比べて，トランザクションの相互依存関係があることと，非線形のトランザクションのおかげで，非同期かつ同時にトランザクションを処理をすることができ，スケラビリティが向上していることがわかる．Sui の Narwhal と Bullshark では，Tangle 構造と block-lattice 構造をより技術的に進めた DAG 構造を構築することができ，概念的にお

ける相違はない．Bullshark と Narwhal プロトコルが Tangle 構造でないのは，ユーザーがトランザクションを行うために他のトランザクションを検証する必要がないからであり，block-lattice 構造でないのは，ブロック検証プロセスで前ラウンドの証明書を必要とするのでウェーブごとに他のバリデーターとのチェーンに依存しているからである．一方で，Bullshark と Narwhal は DAG 構造を応用し，トランザクションが増えるほどネットワークが強固になるという、自己強化の特性を持っている．さらに，特別な承認者がいなくてもネットワークを支えることができる．

### 3.3　DAG を活用するベネフィット

DAG が解決する課題は，(1) 低スループット，(2) 高いレイテンシ，(3) 障害によるパフォーマンスの大幅低下である．

(1) 低スループット．

大量のトランザクションを処理したい時のボトルネックになるのは，ノードの処理能力ではなく，並列計算できないことである．並列計算ができない理由は，blockchain 構造のように一つのチェーンで全てのノードが全てのトランザクションを検証して，ブロックごとに台帳を更新しているからである．

(2) 高いレイテンシ．

多くのブロックチェーンでは，チェーンに書き込む操作は逐次的に行われ，現在のブロックが検証されている間は，トランザクションの流入は停止され，メンプールに格納される．

(3) 障害によるパフォーマンスの大幅低下．

多くのブロックチェーンは障害が起きてもシステムがフリーズすることはないように設計されているが，ほとんど全てのコンセンサスプロトコルで大幅に機能が低下する．例えば，HS はピーク性能で 70,000 tx/s を達成すると上記で示したが，10 ノードのうち，3 つが障害である場合，ピーク性能は 10 分の 1 程度まで低下し，レイテンシは障害なしの 2 秒程度から 15 倍の 30 秒程度まで増加する [22].

これらの課題の克服だけでなく，DAG は以下のベネフィットを享受できる．(1) 並列処理可能による無限のスケラビリティ，(2) ネットワークレイヤーとコンセンサスレイヤーを切り離すことによる通信オーバーヘッドゼロ，(3) マイニングなしなどが挙げられる．

(1) 並列処理可能による無限のスケラビリティ

Sui ではオブジェクト中心設計 [30] により，依存関係ないの共有オブジェクトを並列に処理することができる．他のパフォーマンスが高いブロックチェーンにおいても，例えば，Solana では，Sealevel エンジン [31] を使用して並列処理を行い、Aptos では，Block-STM [32] を使用し，実行時に推論することで並列処理を可能にしている．

(2) ネットワークレイヤーとコンセンサスレイヤーを切り離すことによる通信オーバーヘッドゼロ

これは全ての DAG ベースのコンセンサスで行われているわけではないが，Narwhal とコンセンサスエンジンを組み合わせた手法では，ネットワークレイヤーとコンセンサスレイヤーを切り離すことにより，コンセンサスが追加の通信を必要とせずに完全な順序付けをすることができる．これは，無駄な待ち時間を削減することができ，レイテンシ削減につながる．

(3) マイニングなし

DAG において，ブロックという概念やデータベースを拡張するためにマイニングが必要になるといったことがない．しかしながら，DAG ベースのコンセンサスには，ネットワーク内の全てのバリデータの台帳で同じ状態を保持し，コンセンサスを取るのが困難であり，ほぼ同時に矛盾する二重支払などのトランザクションが書き込まれた場合，これを阻止することができない．これを防ぐためには，ある 1 つのブロック内に矛盾トランザクションを入れないなどの工夫が必要である．

## 4.　Narwhal Mempool Protocol

### 4.1　プロトコル概要

Narwhal メンプールプロトコルの目的は，コンセンサスに提出するトランザクションの可用性（Availability）を満たし，耐障害性を確保しながら，トランザクションの並列順序付けを可能にすることで，大量のトランザクションを同時に処理することができるようにすることや，トランザクションの伝播と順序付けメカニズムを分けることで，システム全体のスループットがコンセンサスのスループットにはほとんど影響されないようにすることである．

**課題.** Tendermint [17], [33] など多くのコンセンサスプロトコルでは，トランザクションの代わりにブロックをブロードキャストして，リーダーはブロックのハッシュを提案しているため，その完全性（Integrity）をメンプールに依存している．つまり，リーダーが提案するブロックの内のトランザクションが正しいかどうかを確認するために，他のノードが自身のメンプールとの内容を照らし合わせて検証しているのである．この結果，ネットワークのパフォーマンスや効率性が損なわれ，低スループットや高レイテンシになる恐れがある．さらに，悪意あるトランザクションによるセキュリティへの悪影響を及ぼすことも考えられる．

**解決策.** Narwhal はトランザクションの順序付けから他のバリデータへのデータ伝播を分離することで，これらの課題を解決している．これは，コンセンサスに提出するトランザクションを一度だけ，直接，ゴシップなしで他のバリデータに分散することで，コンセンサスからの通信オーバーヘッドをゼロにし，低レイテンシに寄与している．また，通信オーバーヘッドがゼロということは，システム全体のスループットがコンセンサスのスループットに影響されづらいということである．Narwhal では，完全な順序づけを必要とするトランザクションを DAG 構造でシーケンスし，その検証をブロックレベルではなくトランザクションレベルで行うことで二重支払問題（Double-spending）を防いでいる．

よって，メンプールとコンセンサスとの分離，コンセンサスで処理するデータの削減，高スループットと低レイテンシのパフォーマンスは，多少分散性においての議論はあるが，Bullshark プロトコルにおいてバリデータノード数増加がネットワークパフォーマンスを向上させることを鑑みれば，PBFT[6] や HotStuff [21] などでは不可能であったパーミッションレスブロックチェーンのトリレンマ問題 [33] の解決策の一つなので

ある．

### 4.2 システムの設計が満たすセキュリティ特性

Narwhal は下記の 5 つの特性を満たしている．完全性（Integrity）とブロックと証明書の可用性（Availability）は，ブロックの拡散とコンセンサスの分離を提供している．包含性（Containment）と 2/3-因果性（2/3-Causality）は，より高いスループットを実現している．1/2-チェーン品質（1/2-Chain Quality）は検閲耐性を提供し，障害の影響を最小化している．

**補題 1: Integrity.** 完全性は，ネットワーク上の不正行為から保護するため，データの処理をするときに常にデータの内容に誤りや欠けが無いことを保証している．honest なノードにより，合意形成に達すると，信頼性が高いデータを取り扱うことが保証され，システム全体の信頼性が向上する．クォーラムインターセクション（Quorum Intersection）を持つプロトコルは，各ブロックが利用可能であり，不正行為がないことを保証している．

**証明.** 完全性は，ハッシュ衝突が起こらないという前提によって導かれる．同じダイジェストに関連付けられた 2 つのブロックを見つけることが不可能であることで，データの完全性が保たれ、改竄やエラーを防ぐことができる．

**補題 2: Availability.** 可用性は，BFT であることを示しており，単一障害点がなく，システムがクラッシュせずに動作し続けることを保証している．Narwhal での可用性は，書き込み操作に対して可用性の証明（certificate of availability）を返すことで確保されている．これは，2f+1 の署名のある投票であり，次のラウンドに含まれる．これにより，任意のブロックがネットワークによって確実に伝播され、使用可能であることが保証されている．

**証明.** 検証者であるバリデーターは自分が署名したすべてのブロックをローカルに保存する．$write(d, b)$ 操作によってブロック b が認証される．そのブロックが認証されるには，$2f+1$ の署名が必要なので，少なくとも f+1 の正直な検証者がダイジェスト d に関連付けられたブロック $b$ を保存している．$read(d)$ 操作では，全ての検証者にリクエストし，$n-f$ の返答を待つ．以前の write 操作が完了した後に呼び出されるどんな write 操作も，少なくとも 1 つの正直な検証者から b を保存しているという返答を得ることができる．

**補題 3: Containment** 包含性は，読み取り操作の際に，$B$ と $B$’ を返すが，$B$’ は $B$ の部分集合である関係の特性のことである．リクエストされたデータが関連データを含むように管理されているため，データの整合性と関連性が保証されている．

**証明.** 各ブロックはそのブロックの作成者とラウンド番号を含む．正直なバリデータは，同じ作成者による同じラウンドで異なる 2 つのブロックに署名することはない．ブロックの認証には $2f+1$ の署名が必要であるから，これらのブロックは認証されることはない．結果，同じ作成者と同じラウンドにおけるブロック作成の一意性が保証される．したがって，正直なバリデーターがローカルに保存する全ての認証済みブロックについて，ブロック内に含まれるダイジェストの集合に包含される前のラウンドからのブロックへの参照は常に一致する．これにより，システム全体のデータ整合性が維持され，情報の矛盾が生じるリスクが低減される．

**補題 4: 2/3-Causality.** 2/3 の因果性は，以前のラウンドからの証明書への参照を含むことで，各ブロックは認証されたかつ利用可能なブロックの因果履歴を保証している．ノードが特定のブロック $B$ の前にブロックの 2/3 を受信することを必要としているので，ブロックの順序付けと因果関係が正確に保持され、ネットワーク全体を同期し，その一貫性が向上する．

**証明.** 各ラウンドに関連付けられたブロックは最大で $3f+1$ 個である．b が証明されたラウンドを $r$ とすると，$B$ は $r$ より小さいラウンドからの最大 $2f+1$ のブロックしか含まないことから，直ちにこの補題が成立する．

**補題 5: 1/2-Chain Quality.** 1/2-チェーン品質は，ビザンチンバリデータによる影響を最小化するため，任意のブロック $B$ が少なくとも半数の正直なノードによって返されたブロックであることを保証している．これにより，不正なノードや攻撃者による操作のリスクが軽減され，ネットワークの安全性と透明性が高まる．1/2-チェーン品質は，検閲耐性を提供する．

**証明.** 各ラウンドにおいてブロックは前のラウンドから少なくとも $2f+1$ のブロックを参照しており，これらの中で最大 f 個がビザンチンバリデーターによって書かれたものであったとしても、残りは正直なバリデーターによって書かれたものであることが保証される．

### 4.3 バリデータの構成要素: Primary-Worker Architecture

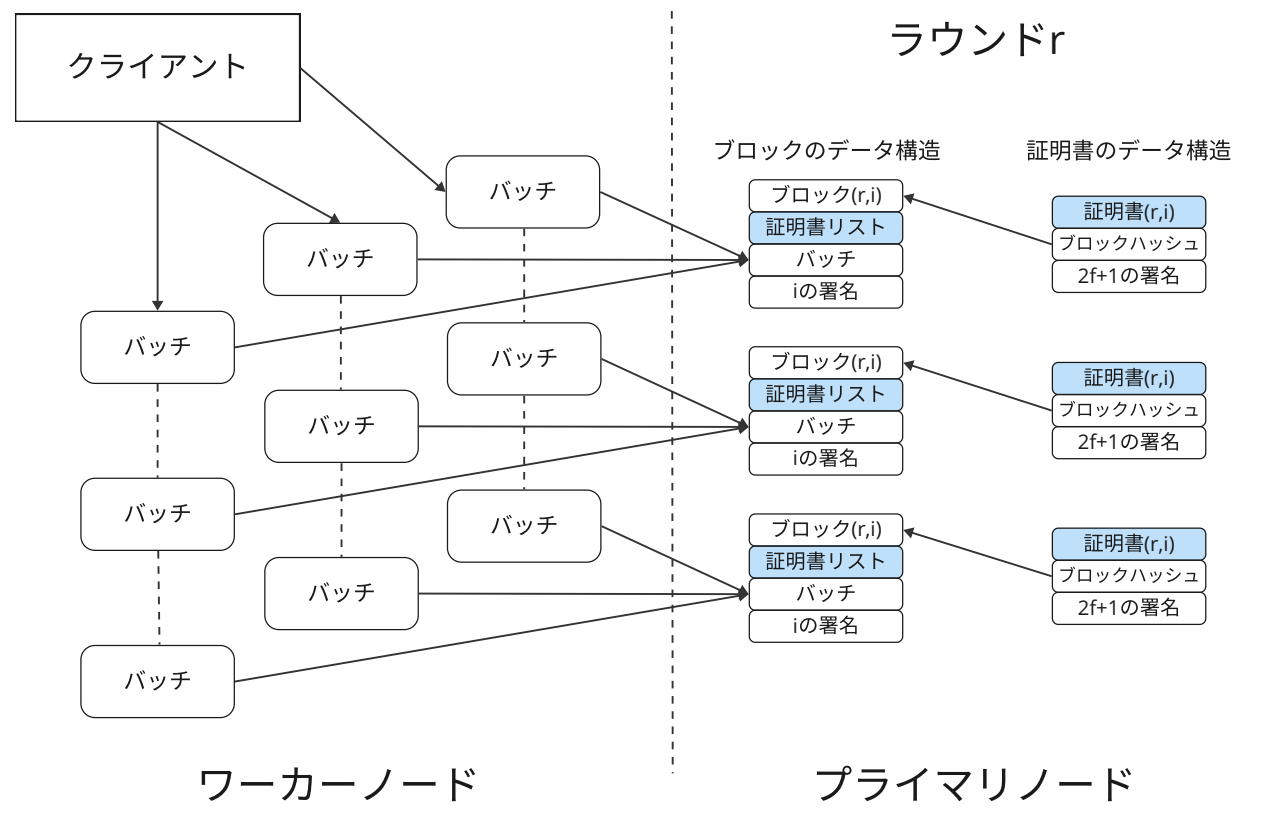

図 4: プライマリワーカーアーキテクチャ．$\forall v \in V$，$\exists w_v \in$ Workers $\wedge \exists! p_v \in$ Primary [注1] の通り，各バリデータ内には DAG 構築のメタデータを取り扱うプライマリノードが 1 台と複数のワーカーノード（例では 3 台）があり，各ワーカーノードがトランザクションバッチを他のバリデータにストリーミングする．バッチはハッシュを次のブロックに含める Primary に送られる．クライアントはすべてのバリデータの Worker にトランザクションを送信する．また，それぞれのノードの初期化やシャットダウンは，Narwhal Manager [注2] によって実行される．

#### 4.3.1 ワーカーノード

ワーカーは，クライアントからのトランザクションの処理と検証（Validation）を行う．また，継続的に他のワーカー同士とデータのバッチを送信し続け，データのバッジは一定蓄積されると，データのバッジのハッシュとしてのダイジェストをバリデータ内のプライマリに転送する．バッジは Sui ではコレクションとも呼ばれる．ワーカー同士がバッチを交換するのは，不要な重複通信を避け，ネットワーク全体でデータが迅速に伝播されることで，効率性や信頼性，特定の種類の攻撃や障害に対する耐性（セキュリティ）を向上するためである．これら一連のワーカーの処理はプライマリが動作中でも関係なくネットワークスピードで動作し続ける．つまり，データの伝播をすることでデータの整合性や順序付けを保つことがワーカーの役割である．

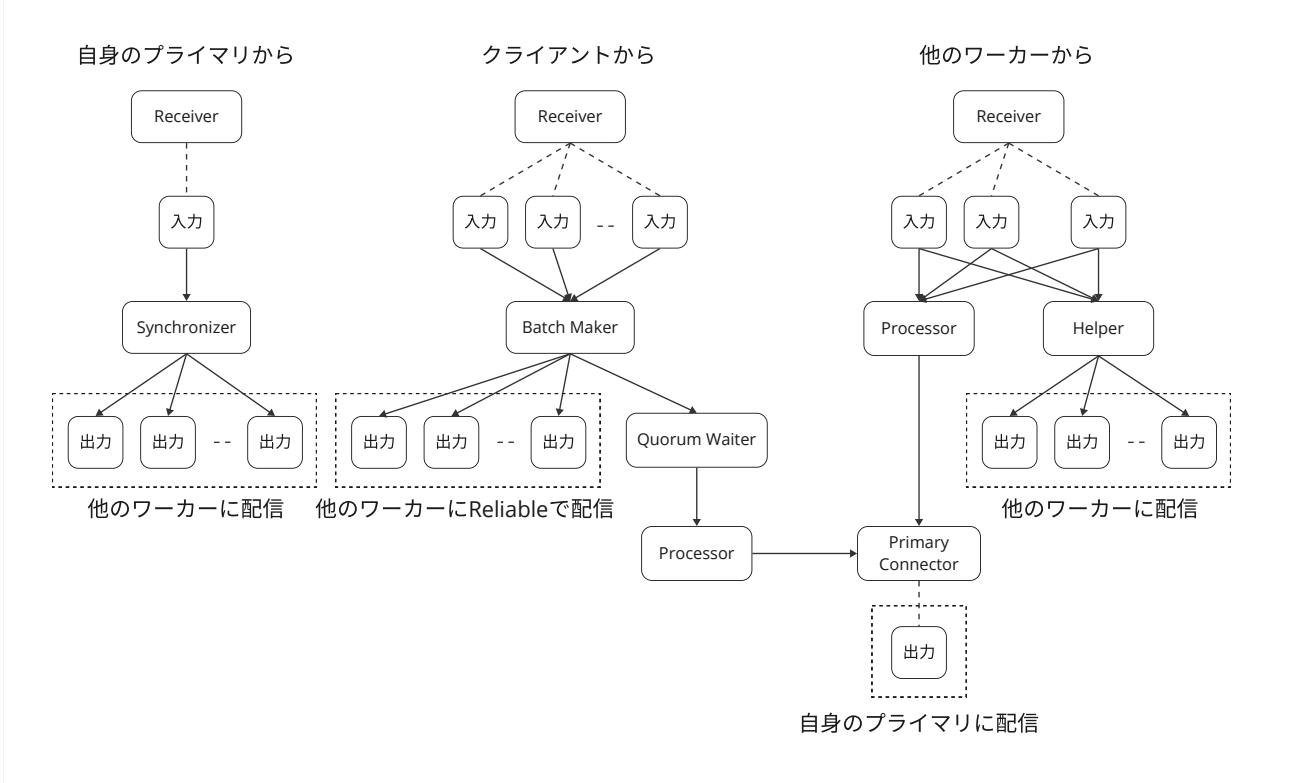

図 5: ワーカーノードアーキテクチャ.(注3)

[20] より，ワーカーノードを使うことで，レイテンシが抑えられることがわかっている．さらに，各バリデータはワーカーを複数持つことによって，容易にスケールアウトすることができる．つまり，各バリデータの割当可能なリソースに応じて，線形的にスケーリングすることができるということである．[20] デフォルトのワーカーは 4 ノード(注4)であるが，$\forall v \in V$，$W(v) = \{w_{v,1}, w_{v,2}, \ldots, w_{v,m}\}$ で示した通り，最大 1000 ノード(注5)まで任意にワーカーのキャパシティを増やすことができる．これにより，バリデータ全体のスループットの向上につながる．

**各コンポーネントとその役割.**

- Batch Maker(注6)

1) Batch Maker は複数のトランザクションをバッチとしてまとめる，2) それをローカル上に保存する，3) 他のバリデータのワーカーノードや自身のプライマリノードへバッチを送信する，4) トランザクションがバッチに含まれた時点で，バッチのダイジェストをトランザクション送信者であるエンドユーザにフィードバックするという役割を担っている．

- Quorum Waiter(注7)

Quorum Waiter は新しいバッチがこのチャネルを通じて受信されると，他のワーカーへバッチをブロードキャストする．送信後，他のワーカーからの ACK を待ち，ACK を送り返されることによって，バッチが正しく受信され，検証されたことが確認できる．Quorum を達成するためにはワーカーごとのステーク量によって重みが算出され，必要なステーク量をを持つ ACK を得る必要がある．Quorum が達成されると，バッチは確定され，そのダイジェストがプライマリに送信される．この時点で，そのバッチはネットワークによって承認されたものであるとみなすことができ，次の処理へ進むことができる．

- Processor

Processor はバッチのダイジェストを作り，それをローカルに保存する．

- PrimaryConnector

PrimaryConnector はバッチのダイジェストをプライマリに送信する．

- Synchronizer

Synchronizer は，自分のプライマリに対して必要なバッチをリクエストする

- Helper

Helper は，他のワーカーに対して必要なバッチを返信する

- Receiver

Receiver は必要なデータを受け取り，ワーカーが何らかのデータを受信するときこれが動作する．この中でも，TxReceiverHandler(注8)がクライアントからトランザクションを受け取っている．

#### 4.3.2 プライマリノード

プライマリノードはバリデータ内の中心的な役割を果たすマシンであり，Bullshark コンセンサスに渡すラウンドベースの DAG を構築とその準備や調整を担う．プライマリノードはデータの伝搬速度に制約されないが，ワーカーノードは任意の数だけ増やすことができるので，処理のボトルネックは概ねプライマリとなる．

**各コンポーネントとその役割.**

（注1）：Mysten Labs, ”Narwhal manager”, GitHub, https://github.com/YuseiWhite/sui/blob/mainnet/crates/sui-core/src/consensus_manager/narwhal_manager.rs, February 2025.

（注2）：Mysten Labs, ”Narwhal manager”, GitHub, https://github.com/YuseiWhite/sui/blob/mainnet/crates/sui-core/src/consensus_manager/narwhal_manager.rs, February 2025.

（注3）：Mysten Labs, ”Worker diagram,” GitHub, https://github.com/YuseiWhite/sui/blob/mainnet/narwhal/worker/README.md, February 2025.

（注4）：Mysten Labs, ”Workers,” GitHub, https://github.com/YuseiWhite/sui/blob/mainnet/narwhal/Docker/validators/workers.json, February 2025.

（注5）：Mysten Labs and Facebook, ”Channel capacity”, GitHub, https://github.com/YuseiWhite/sui/blob/mainnet/narwhal/worker/src/worker.rs#L49, February 2025.

（注6）：Mysten Labs and Facebook, ”Batch maker,” GitHub, https://github.com/YuseiWhite/sui/blob/mainnet/narwhal/worker/src/batch_maker.rs, February 2025.

（注7）：Mysten Labs and Facebook, ”QuorumWaiter::run function”, GitHub, https://github.com/YuseiWhite/sui/blob/mainnet/narwhal/worker/src/quorum_waiter.rs L85-L187, February 2025.

（注8）：Mysten Labs, ”Transaction receiver handler struct,” GitHub, https://github.com/YuseiWhite/sui/blob/mainnet/narwhal/worker/src/transactions_server.rs#L132-L136, February 2025.

- Proposer(注9)

Proposer は，ヘッダーに関する処理を担い，十分な数のダイジェスト（デフォルトでは 32(注10)）と親証明書がある場合，新たにヘッダーを生成する．ヘッダー生成の最大遅延のデフォルトは 1,000ms (= 1 秒)(注11)である．さらに，作成されたヘッダーを全てのバリデータにブロードキャストし，それを証明するために，Certificater(注12)にも送信する．

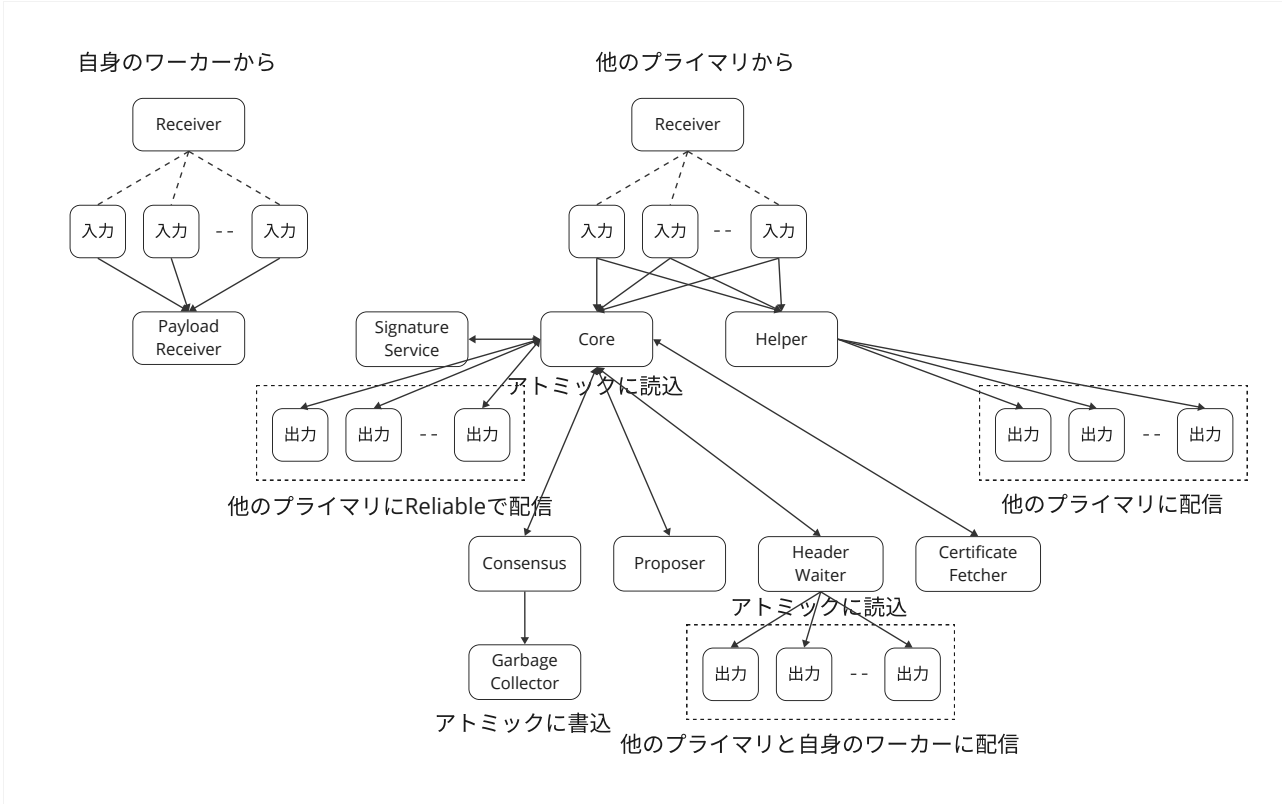

図 6: プライマリノードアーキテクチャ.(注13)

- Consensus(注14)

Consensus では，新たに Bullshark 構造体を作り，コンセンサスを実行し，さらにクライアントのトランザクションの実行の取り扱いまでを行う．

- Header Waiter

Header Waiter は他のプライマリや自身のワーカーに対して，バッチと証明書を要求する．

- Garbage Collector

Garbage Collector はガベージコレクションのラウンドを更新する．デフォルトは 50 ラウンドである．(注15)

- Certificate Fetcher

Certificate Fetcher は証明書の全ての履歴を待つ．

**Primary-Worker Architecture における通信とその関数**

Worker to Worker

- report_batch() 関数

ワーカーから受信したバッチを検証して，ワーカーの ID とここで計算したバッチのダイジェストを送信する．

- request_batches() 関数

ワーカーから一意の特定のバッチのダイジェストをもとにそのバッチを要求する，バッチのリストを返す．

Primary to Worker

- synchronize() 関数

プライマリから特定のバッチのダイジェストのリストを受け取り，足りないバッチを他のワーカーに同期してもらい，そのバッチを検証しローカルに保存する．

- fetch_batches() 関数

プライマリから特定のバッチのダイジェストのリストを取得し，それのバッチをワーカーから受け取り（同期），バッチのリストを完成させて返す．Primary to Worker

Worker to Primary

- report_own_batch() 関数

ワーカー自身が処理したバッチのダイジェストとワーカー ID などのバッチ関連情報をプライマリに送信し，同期されたことを確認する．この関数により，プライマリはトランザクションの進行状況を確認でき，そのバッチから新しいヘッダーを提案する．report_others_batch() でも同様である．

- report_others_batch() 関数

他のワーカーからバッチのダイジェストやワーカー ID などのバッチ関連情報を受け取り（同期），ローカルに保存する．

Primary to Primary

- send_certificate() 関数

新しい証明書が作成されると，その証明書をプライマリに送信（同期）する．

- send_randomness_partial_signatures() 関数

新しいヘッダーが提案されると，他のプライマリに検証と投票をしてもらう要求を行う．

- request_vote() 関数

特定のラウンド，Authority ノードの身元に基づいて他のプライマリに証明書を要求（同期）する．

- fetch_certificates() 関数

ランダムネスの部分署名を要求する Authority ノードの身元を確認し，その部分署名を他のプライマリに送信する．

## 5. Bullshark Consensus Protocol

### 5.1 概　　要

Bullshark は，Narwhal メンプールの DAG 上に構築された，ゼロオーバーヘッドの BFT アプローチのビザンチンアトミックブロードキャストプロトコルである．部分同期バージョンの最適化を実用化できたことにより，ガベージコレクションと公平性の競合を解決した初めてのプロトコルである．これにより DAG-Rider [34] の望ましい特性を全て維持しながら同期期間中のレイテンシを削減することができる．また，同期期間中で発生するビューチェンジとビュー同期という複雑なプロセスを完全に排除できたことにより，大幅なレイテンシが起きることは滅多に無い．

### 5.2 技術的課題

#### 5.2.1 DAG ベースコンセンサスプロトコルにおける課題

DAG ベースのコンセンサスプロトコルは，前述した通り，低

(注9)：Mysten Labs and Facebook, "Proposer," GitHub, https://github.com/YuseiWhite/sui/blob/mainnet/narwhal/primary/src/proposer.rs, February 2025.

(注10)：Mysten Labs, "Batch size," GitHub, https://github.com/YuseiWhite/sui/blob/mainnet/narwhal/Docker/validators/parameters.json#L4, February 2025.

(注11)：Mysten Labs, "Max header delay," GitHub, https://github.com/YuseiWhite/sui/blob/mainnet/narwhal/Docker/validators/parameters.json#L8, February 2025.

(注12)：Mysten Labs, "Certifier struct," GitHub, https://github.com/YuseiWhite/sui/blob/mainnet/narwhal/primary/src/certifier.rs#L40-L68, February 2025.

(注14)：Mysten Labs, "PrimaryNodeInner::spawn_consensus", GitHub, https://github.com/YuseiWhite/sui/blob/mainnet/narwhal/node/src/primary_node.rs#L286-L379, February 2025.

(注15)：Mysten Labs, "GC depth," GitHub, https://github.com/YuseiWhite/sui/blob/mainnet/narwhal/Docker/validators/parameters.json#L3, February 2025.

スループット，高いレイテンシ，障害によるパフォーマンスの大幅低下を解決する仕組みであるが，どのように DAG を構築するか，どのようにブロックチェーンに書き込む準備をするかというプロセスの中で問題が発生する場合がある．一貫性（Consistency）を蔑ろにし，可用性（Availability）を重視した DAG ベースのコンセンサスプロトコルも多い．例えば，IOTA [35] は Tangle 構造を使用しており，これは 1 つのトランザクションが親となる 2 つのトランザクションを参照し，それを繰り返すことによって非構造化で DAG を構築している．各トランザクションが過去の複数のトランザクションを承認する仕組みによりトランザクションデータの改竄を防ぐことができているものの，半順序（Patial Order）であるため一貫性を保証できない．つまり，すべてのバリデータが合意するトランザクションの一貫した順序を確立されないため，複数の競合するトランザクションがブロックチェーンに追加される可能性があり，悪意のあるアクターにより競合するトランザクションが送信される恐れがある．また，一貫性を保証できないことは確率論的ファイナリティしか達成できず，理論的に二重支払問題を解決できない．DAG ベースコンセンサスプロトコルの他の研究 [6], [36] において，可用性を重視したプロトコルのほとんどは，たとえ構造化された DAG だとしても，確率論的ファイナリティしか達成できないことが示されている．HashGraph [37] や Cordial Miners [38] など Optimistic な DAG を採用する一貫性を重視したプロトコルにおいても，Best-effort（BE）でメッセージをブロードキャストしている．これは，通信オーバーヘッドを減らすことができるものの，確実なメッセージ伝播が保証できないため，一貫性に必要な合意形成のメッセージを受信できない可能性があり，結果的にも理論的にも一貫性を損なう恐れがある．また，DAG の構造上の問題として，並列処理可能な DAG は，最終的にトランザクションまたはブロックのシーケンスを行う際の計算コストが多くなり，一意のシーケンスを行うためのプロトコルが複雑となり，その実装と検証が困難となる．バグや脆弱性が生じるリスクも増大する．

#### 5.2.2 理論的な DAG-Rider における課題

Bullshark プロトコルも，バリデータ間で証明書によって認証した構造化された DAG を使用しているが，これは Reliable Broadcast Protocol によって達成されている．Reliable なブロードキャストは BE よりも多くの通信プロセスを経るために，通信のオーバーヘッドが発生する．通信のオーバーヘッドが引き起こされるとレイテンシも増加する．コンセンサスレイヤーへのサブ DAG への送信は Narwhal のプライマリノードが担っている．ここでコンセンサスレイヤーはネットワークの状態によって，追加の署名や証明書など必要なデータがある場合は，バリデータにリクエストする必要がある．これはビザンチン環境や順序付けするまでのプロセスで起こり，これが理想的な TPS を阻害し大幅なレイテンシをもたらす．Bullshark の理論的な DAG-Rider [34] は非同期ネットワークを仮定しているため，DAG ベースでない Tendermint [17], [33] や HotStuff [21] よりもレイテンシが増大する場合がある．また，実用面で言えば，公平性を維持するための無制限のメモリが必要ということは非現実的な仮定である．

#### 5.2.3 BAB 問 題

Reliable Broadcast Protocol は，システム内に一定の faulty なノードが存在しても reliable なメッセージの伝播を保証することによってシステムの信頼性を向上させ，Agreement, Integrity, Validity を満たすことができるが，全順序（Total Order）を保証していない．これは，二重支払を可能とし，トランザクションに関連するデータの整合性や一貫性も保証できないため，DLT というシステムの信頼性を全て失うことにつながる．Reliable Broadcast Protocol が満たす Agreement, Integrity, Validity だけでなく全順序，Total Order も満たす Byzantine Atomic Broadcast（BAB）問題を取り上げる．関連研究 [39], [40] によれば，BAB 問題とは，すべてのプロセスに同じメッセージを同じ順序で配信する問題のことである．

**Validity.** 有効性とは，正しいプロセスがメッセージ M をブロードキャストした場合，1 つ以上の正しいプロセスが最終的に M を配信していることである．

**Agreement.** 合意とは，正しいプロセスがメッセージ M を配信した場合，すべての正しいプロセスは最終的に M を配信していることである．

**Integrity.** 完全性とは，任意の識別子 ID に対して，すべての正しいプロセスは識別子 ID を持つメッセージ M を最大 1 回だけ配信し，sender(M) が正しい場合，M はすでに sender(M) によってブロードキャストされていることである．

**Total Order.** 全順序とは，2 つの正しいプロセスが 2 つのメッセージ M1 と M2 を配信する場合，両方のプロセスは 2 つのメッセージを同じ順序で配信していることである．複数のワーカーノードはワーカーノードの集合として記述する．

Byzantine Atomic Broadcast Protocol は，ほとんどのビザンチン SMR システムで実現されているシーケンシングプロトコルよりも強い保証を提供している．DAG-Rider [34] は非同期設定でこれを満たしているが，FLP 不可能性 [41] により，非同期通信において，たとえ一つのプロセスが故障する可能性があるだけであっても，全てのプロセスが常にコンセンサスに達する決定論的なアルゴリズムは存在しないことが証明されている．よって，実用面で Byzantine Atomic Broadcast Protocol を満たすためには，同期通信または部分同期の設定で，アルゴリズムの設計が必要となる．

#### 5.2.4 公平性とガベージコレクションのトレードオフ

公平性とは，BAB 問題の Validity を満たすことと同義である．DAG-Rider [34] は確かに公平性を満たしているものの，FLP 不可能性 [41] により，非同期設定において，ブロックをブロードキャストしない faulty なパーティと古いラウンドをガベージコレクションする前に，待つ必要のある遅いパーティを区別できないことから，対象にしたいパーティだけをガベージコレクションを行うことができない．前のラウンドでまだ順序付けされていないブロックを参照する弱いリンクを利用することによって，すべてのブロックが最終的に順序付けされることを保証しているが，これも理論的なものでどのようにガベージコレクションを行うか示されていない．

一方．Narwhal はガベージコレクションメカニズムを実装しているものの，BAB 問題の公平性を犠牲にしている．その理由は，遅いパーティのブロックは全順序付けされる機会を得る前にガベージコレクトされる可能性があるため，すべての当事者に公平性を提供していないところにある．

## 5.3 Bullshark Consensus Protocol

### 5.3.1 Bullshark の特徴

**ネットワーク通信レイヤーとコンセンサスレイヤーの分離.** Bullshark コンセンサスレイヤーは，Narwhal でのネットワーク通信レイヤーとを切り離して実行している．これは，Reliable Broadcast Protocol による通信オーバーヘッドの可能性をコンセンサスレイヤーに持ち込まないということである．ネットワーク通信レイヤーとの分離は，ネットワークの状態を無視しているため，通信オーバヘッドゼロでコンセンサスロジックを進めることができ，また，抽象化することができるようになるため，シンプルで効率的な実装・保守が可能となる．ネットワークの状態を無視しているというのは，トランザクションの順序付けというメタデータの処理と実際のトランザクションのデータ配信をそれぞれ独立して行うことができていることである．さらに，この分離によってネットワークの変更や最適化も容易である．BullShark の順序付けロジックは，DAG を構築する上で通信を必要とせず，各パーティは DAG のローカルコピーを検証し、エッジを投票として解釈することで頂点を全順序付けしている．パフォーマンスにおいてもスケーラブルで高スループットのコンセンサスシステムを可能としている．しかし，セキュリティ面では，切り離しの方法や，データの検証と認証の方法に依存するため，切り離せばそのままコンセンサスと実行を行えるわけではないことを留意されたい．

**ラウンドベース DAG.** ラウンドベース DAG は Aleph [42] で初めて導入された．ラウンドベース DAG とは，$f = Byzantine$，$n = Validators$ とし，頂点はラウンド番号と関連付けられているとき，各バリデータはラウンドごとに 1 つのメッセージをブロードキャストし， 各メッセージは少なくとも異なるバリデータから $n - f$ のメッセージを参照する．ラウンド $r$ に進むには、バリデータはまず，ラウンド $n - f$ の異なるバリデータからメッセージを $r - 1$ で取得する必要があるということである．

Bullshark はあらかじめ定義された方法で偶数ラウンドごとにアンカーリーダーを選出する．Bullshark は単にこのリーダーブロックが結果的にコミットされるかスキップされるかを決めるルールである．ノードがリーダーブロックをコミットすると，そのブロックの因果履歴である，まだコミットされていないすべてのブロックをコミットすることとなる．

また，コミットするにあたって，決定論的ルールにより Bullshark はビューチェンジやビュー同期メカニズムを必要とせず，DAG にエンコードされた情報を使って安全性（safety）を維持している．

### 5.3.2 ラウンドベース DAG コミットプロセス

図 7 では，$n = 4, b = 1$ での Bullshark プロトコルのコミットプロセスの状態を表している．DAG の各奇数ラウンドでは，あらかじめ定義され，選出された Anchor Leader が存在し，図では赤色のブロックで示されている．ここでは，どのアンカーをコミットするかをバリデータが決定することが目的である．DAG の全てのノードを全順序付けする（Total Ordering）ために，各バリデータは，全てのアンカーを確認し，因果関係の履歴を決定論的なルールに従って順序付けを行う．$A2$ の因果関係の履歴は，赤色の枠線で示されている．

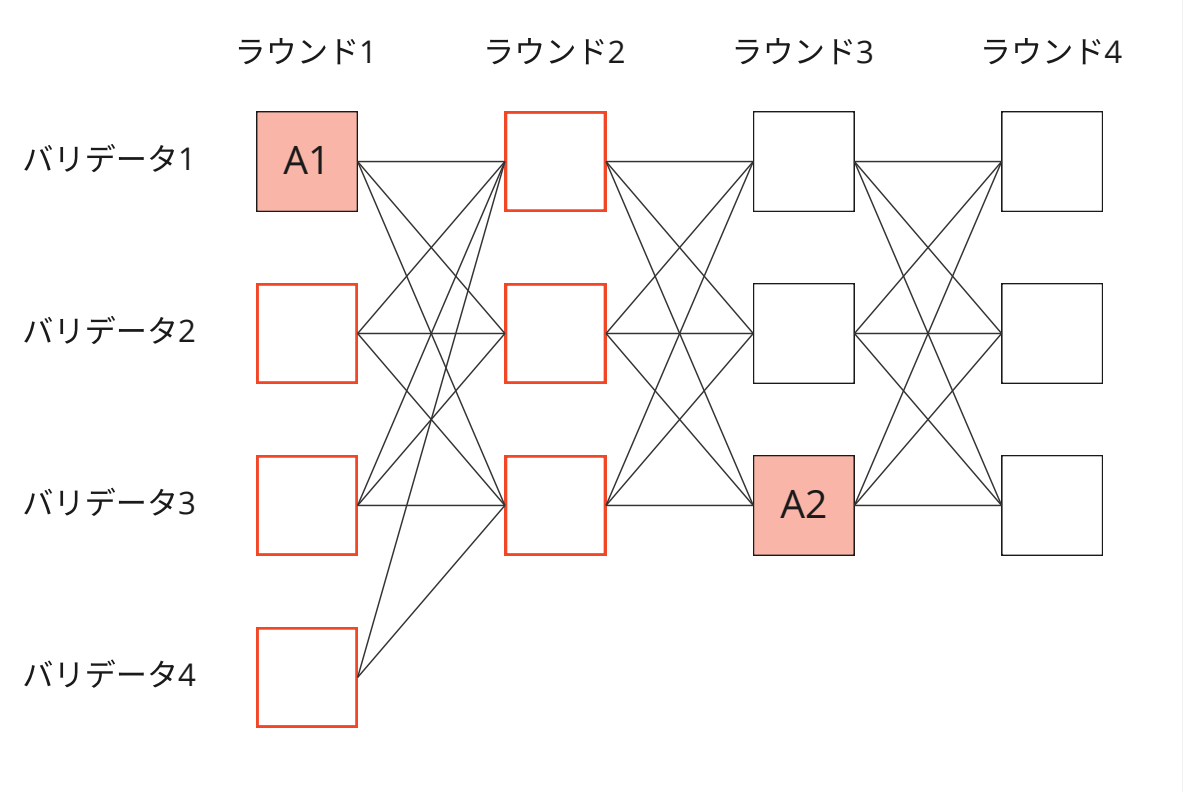

図 7: 4 台のバリデータにおける DAG コミットプロセス．

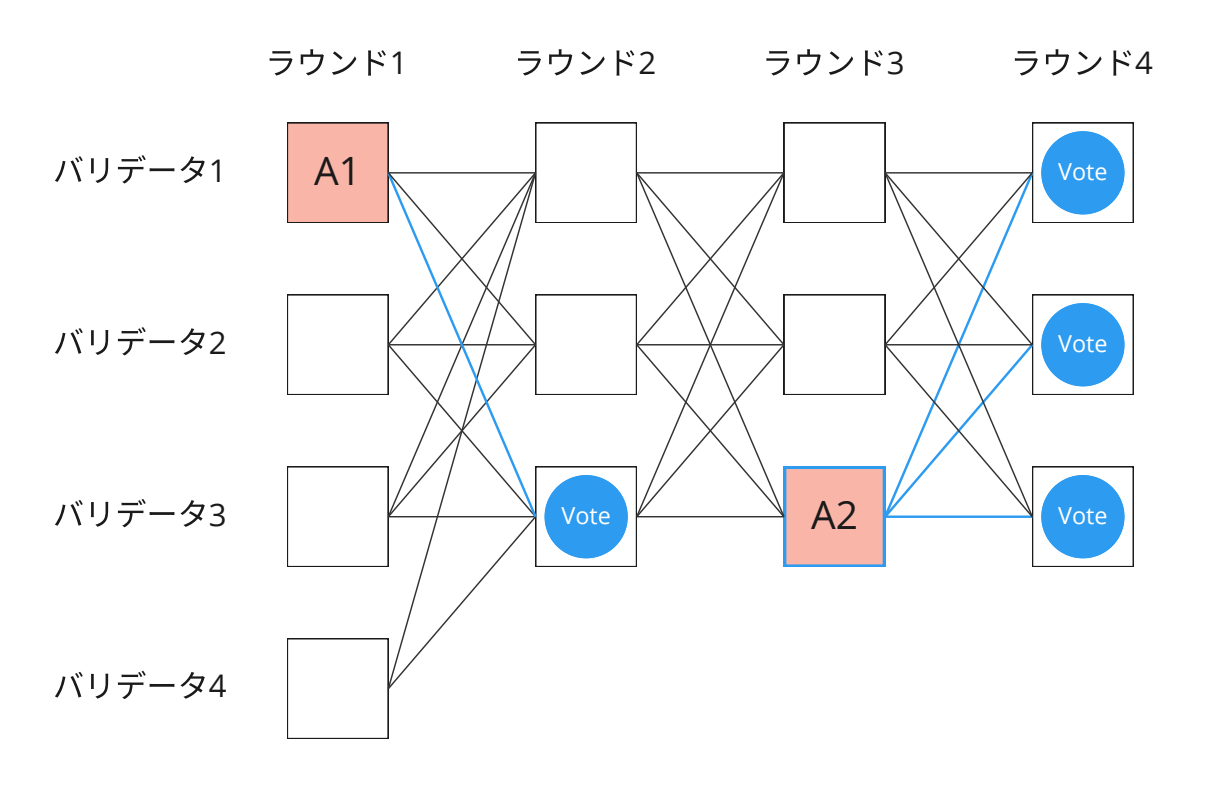

図 8: 4 台のバリデータにおける投票プロセス，バリデータ 1 のローカルビュー

図 8 では，各バリデータが Anchor leader に投票し，DAG がコミットされる状態を示している．偶数ラウンドでは，各バリデータが前ラウンド $(r - 1)$ の Anchor leader に対して 1 票を投じることができる．コミットルールはシンプルで，アンカーは $f + 1$ 以上の票を獲得すればコミットされ，コミットされたトランザクションは決定論的ファイナリティを達成する．トランザクションがシステムに追加されるには，その前のトランザクションが参照されている必要がある．図においては，A2 は 3 票を獲得して，3 つのエッジがありコミットされるが，A1 は 1 票しか獲得しておらず，$f + 1$ を達成できないためコミットされない．また，投票は技術的には単なる署名である．ビザンチンフォルトトレランス性を満たすのが $f + 1$ であるというのは，少なくとも 1 つは honest なバリデータからのものであることが保証されていることを示しており，f は遅延と悪意のある投票しか含まれていないことにある．つまり，悪意のあるトランザクションはすでに Narwhal プロトコルが DAG を構築するまで

に省かれており，Bullshark プロトコルではトランザクションの検証（validation）はほとんど不要である．

図 9 では，バリデータ 1 とバリデータ 2 のローカルビューの違いとアンカーが順序付けされていることが示されている．図 7 をバリデータ 1 のローカルビューとした時，バリデータ 1 は A1 に対して Quorum を達成していないが，図 8 のバリデータ 2 は A1 に 2 票（$f+1$）が投じられているため，A1 はコミットされる．

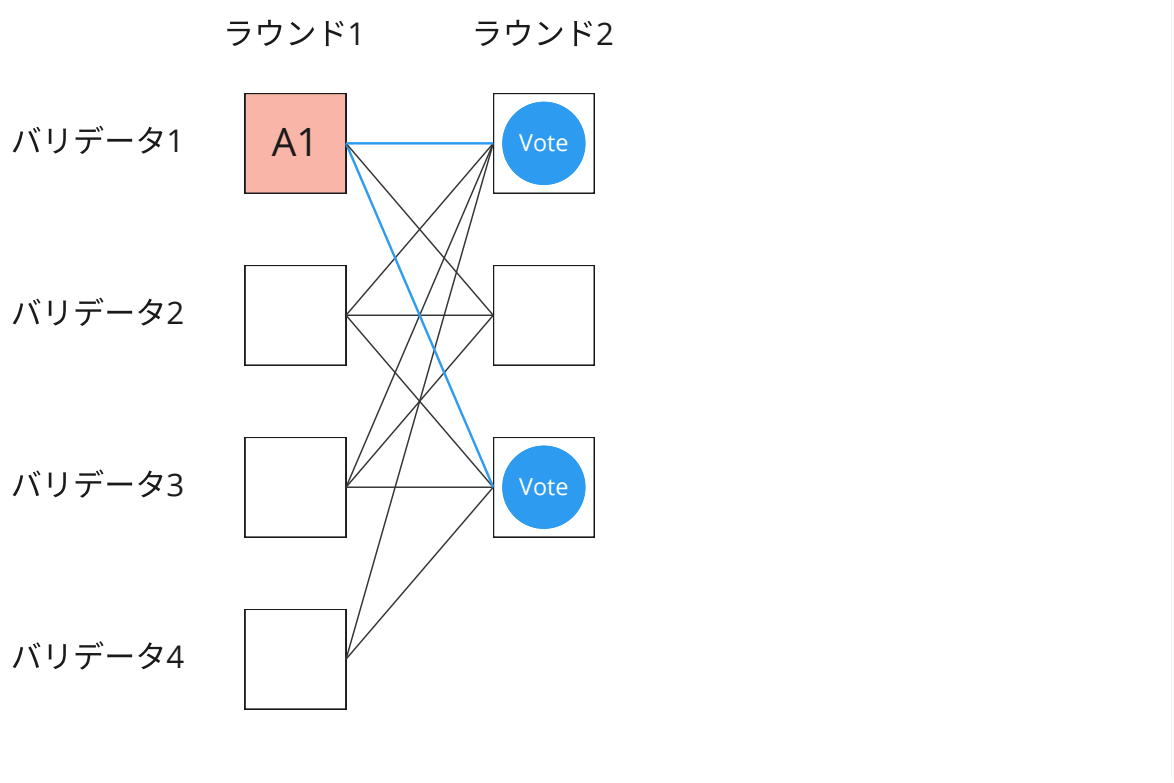

図 9: 図 6 に対してのバリデータ 2 のローカルビュー

全順序を保証するためには，バリデータ 1 は A1 を A2 より前で順序付けしなければならない．これは Quorum Intersection によって達成される．Quorum Intersection は，任意の 2 つの Quorum が少なくとも 1 つの honest なノードを共有することを保証する性質であり，Quorum とは，意思決定に必要な最小限の集合である．Quorum Intersection を用いることで，ブロックチェーンが決定論的な方法で拡張されることやトランザクションがアトミックにコミットされることを実現することができる．Quorum は $2f+1$ のバリデータであり，任意の 2 つの Quorum は，少なくとも 1 つの honest なノードを共有でき，この honest なノードは，両方の Quorum に正確な情報があるという異なるバリデータ間の情報の一貫性を保証し，異なる Quorum の矛盾する決定がある場合の解となる．システム全体で $3f+1$ のノードがあるので，$(2f+1)+(2f+1)-(3f+1)=f+1$ より，$f+1$ の投票で少なくとも 1 つの honest なノードが存在することを保証することができる．バリデータ 2 が A1 に対して $f+1$ の投票があることを観測した場合，Quorum Intersection により，他のバリデータにもこの情報が伝播し，バリデータ 1 が自身のローカルビューで A1 に対する $f+1$ の投票を持っていなくても，この情報と DAG の構造，各ノードが親ノードの $n-f$ の署名を持っていることから，自身のローカルビューを更新し A1 を A2 より前に順序付けすることができる．A1 以前のアンカーが存在しても最後にコミットされたリーダーまで再帰的に順序づけられているかをすでに確認しているのでここで正しく順序づけられているかを確認する必要はない．

よって，safe-to-skip の定理を適用することができる．これは，特定の条件下でアンカーをスキップしても安全であることである．あるアンカーが将来のアンカーからリンクがない場合，そのアンカーはどのバリデータにも投票されていないことが保証されるため，安全にスキップできる．これにより，faulty なリーダーによるコミットプロセスの遅延を防ぐことができる．

### 5.3.3 リーダータイプとリーダー選出

Bullshark では，各ウェーブごとに 2 つの種類，3 つのリーダーを持っている．Steady-State Leader と Fallback Leader が 2 つのリーダータイプである．Steady-State Leader は，同期期間中に各ウェーブごとにコミットされる，先述のアンカーリーダーと同様である．1 ウェーブは 4 ラウンドであるため，ウェーブごとに 2 つの Steady-State Leader が選出される．Bullshark では，外部のビュー変更やビュー同期メカニズムが不要であるため，ラウンド 1 のリーダーが honest であればラウンド 3 をほとんど同時に開始することができる．Fallback Leader は，非同期期間中に 2/3 の確率，平均 6 ラウンドでコミットされる．これは liveness を確実に保証するための特殊なケースのリーダーであり，Tusk[18] と同様である．Quorum Intersection により，異なるリーダータイプが同じウェーブでコミットされることはない．よって，頂点 $v$ をバリデータ $v_i$ が配信し，$v$ の因果履歴がリーダーをコミットするに十分な情報がある場合は，他のバリデータ $v_j$ は $v_i$ の投票タイプを Steady-State Leader とし，そうでない場合は Fallback leader とする．また，Reliable Broadcast Protocol のおかげで全てのバリデータは頂点は，$v$ の同じ因果履歴を見ることができるので，同じ投票タイプを合意, agree できる．ビザンチンであろうとこのリーダータイプを欺くことはできない．

リーダー選出は，Steady-State Leader は奇数ラウンドでブロードキャストするために，偶数ラウンドに LeaderSchedule::leader() 関数(注16)でプロダクション環境ではラウンドでシードされたステーク加重方式でリーダーを選出する．Fallback Leader は，ラウンド 1 で配置するために，ラウンド 4 で生成されたランダム性を使用して遡及的に選出される．

### 5.3.4 公平性の達成とガベージコレクションの実装

DAG-Rider [34] の公平性メカニズムと Narwhal [20] のガベージコレクションメカニズムは競合する．よって，これに対して，Bullshark はガベージコレクションを実装しながらも DAG-Rider [34] で理論化された弱いリンクを効果的に活用し，互いの利点をうまく組み合わせ，同期期間中のみ公平性を保証する．

Hashgraph [37] では DAG が構造化されていないため，新しいブロックの validity を検証するために DAG のプレフィックス全体をメモリに保持しなければならない．これを実現するためには無限のメモリが必要であるため実用レベルで実装することはできない．Narwhal [20]，DAG-Rider [34]，Aleph [42] では，ラウンドベースで構造化された DAG を使っているが，公平性を犠牲にしている．非同期ネットワークにおいてメモリを実装する際，全ての honest なバリデータに対しての公平性を提供することは難しいため，Bullshark では，Global Stabilization Time

（注16）：Mysten Labs, "LeaderSchedule::leader function," GitHub, https://github.com/YuseiWhite/sui/blob/mainnet/narwhal/primary/src/consensus/leader_schedule.rs#L233-L263, February 2025.

(GST）の後，honest なバリデータによる全てのメッセージを有限時間でガベージコレクションの前に DAG へ入れている．GST の後には ◇$P$ Failure Detector [43], [44] が必要である．

◇$P$ Failure Detector は，全ての faulty なプロセスは，最終的に faulty でないプロセスによって永続的に fault であると疑われるようになる Strong Completeness と最後のクラッシュが発生した時点を初期の不安定期の終了と定義し，その後ある程度の時間が経過すると，non-faulty プロセスが誤って障害と判断されることはなくなる Eventual Strong Accuracy によって，全てのバリデータから頂点を得ることができていなくてもラウンドをガベージコレクトすることができる．これを Bullshark は以下のタイムスタンプの概念を実装することで実現している．Bullshark は有界メモリを維持しながら，正しいプロセスがメッセージ $M$ をブロードキャストした場合，GST の後，全ての正しいプロセスが最終的に $M$ を配信していることを保証する．ガベージコレクションメカニズムのために，全ての頂点，vertex に対してタイムスタンプを追加し，honest なバリデータは頂点 $v$ をブロードキャストした時の時間 $v.ts$ を付加する．そしてバリデータはガベージコレクションのラウンド数分（デフォルトは 50 ラウンド(注17)）を維持し，それ以上のラウンド数の DAG には頂点を追加せずにメモリを解放する．

ブロックチェーンのコンセンサスプロトコルにおける公平性は，どんなバリデータも計算効率に関係なく，たとえ遅延したバリデータでもトランザクションのコンセンサスに貢献することができるということであり，DAG ベースでは全てのバリデータが DAG に頂点を追加できるようにすることがである．Bullshark は同期期間中の DAG 構築にタイムアウトを導入し弱いリンクを提供する．弱いリンクは，通常のリンク（アンカーリーダー間が連結されている状態）と違って，最新ラウンドから 2 つ以上前のラウンドに対して faulty でないバリデータは他で Quorum を達成していないアンカリーダーに投票でき，そのリーダーが通常と同条件でコミット対象とするということである．これにより，遅延したバリデータもコンセンサスに貢献することができ，同期期間中において公平性が保証される．

## 6. ブロックチェーンコミットまでのアルゴリズム

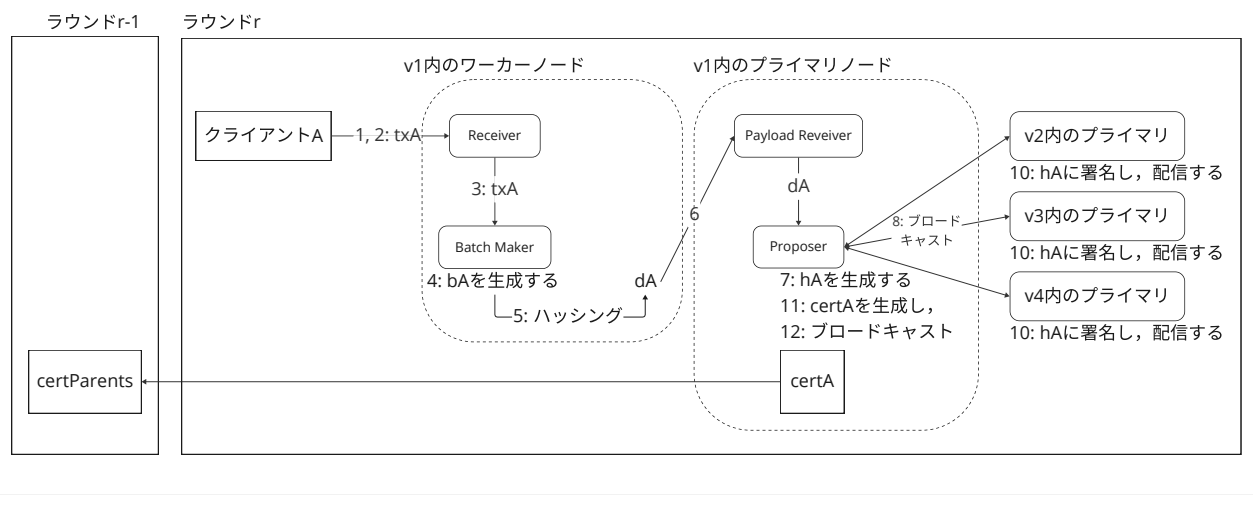

図 10: Narwhal によるラウンド DAG 構築アルゴリズム．

［定義］

$V = \{v_1, v_2, \ldots, v_n\}$

V はバリデータ $v_i$ の集合（$Validators$）を指し，$v_i$ は単一の validator を指す．

$\forall v \in V$，$\exists w_v \in \text{Workers} \land \exists! p_v \in \text{Primary}$ [63]

各バリデータはそれぞれ複数のワーカーノードと単一のプライマリノードを構成している．

$\forall v \in V$，$W(v) = \{w_{v,1}, w_{v,2}, \ldots, w_{v,m}\}$，$default: m = 4$

複数のワーカーノードはワーカーノードの集合として記述する．

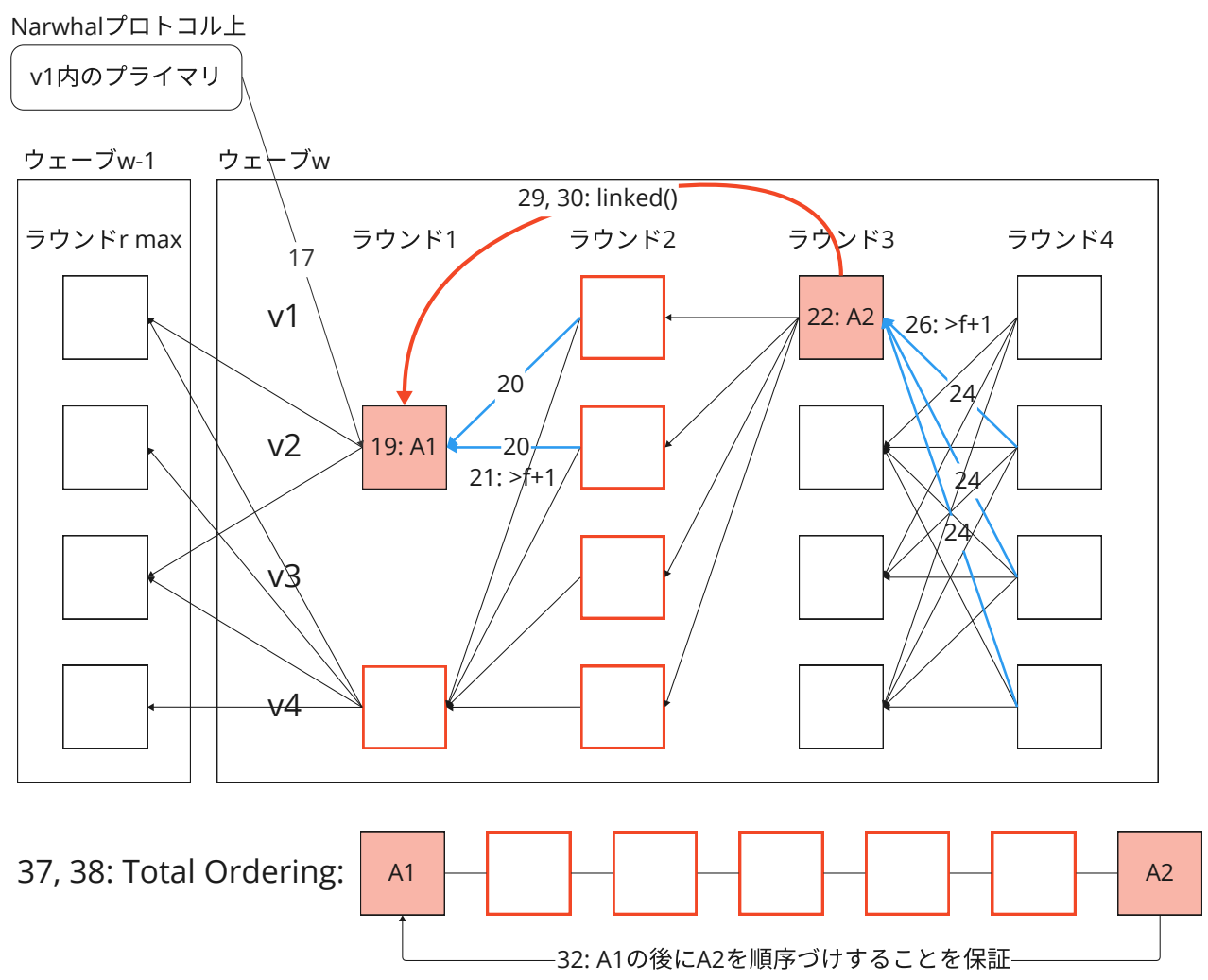

図 11: Bullshark による Total Ordering アルゴリズム．

**Network レイヤー: Narwhal によるラウンド DAG 構築．**

（1） クライアントユーザ $client_A$ はバリデータ $v_1$ にトランザクション $tx_A$ を送信する．[66] 最小トランザクションサイズは 9 バイト [44] である．クライアントユーザは $V = \{v_i \mid i \in \mathbb{N}, 1 \leq i \leq n\}$ に対してトランザクションや詳細を後述する証明書をマルチキャストでき，この作業の一部または全てをサードパーティのサービスプロバイダに任せることができる．このサービスプロバイダは safety は不要であるが liveness は必要である．[45]

（2） ワーカーノード $w_{1,1} \in v_1$ が $tx_A$ を受信する．

（3） $w_{1,1}$ は $tx_A$ を validate() 関数(注18)を実行することで検証し，有効である場合，$tx_A$ はバッチに追加される．

（4） $batchMaker \in w_{1,1}$ は $tx_A$ や他のトランザクション $Tx = \{tx_b, tx_c, \ldots, tx_k\}$，$tx_i \in \mathbb{N}$，$b \leq i \leq k$ をまとめたバッチ $b_A$ を作成する．また，$w_{1,1}$ は $\{v \in V \mid v \neq v_1\}, \exists w \in v$ に対して $b_A$ を送信し，受信した各ワーカーノードは $b_A$ をローカルで格納する．

（5） $b_A$ はワーカーノード内でハッシュ化され，ダイジェストが計算される．$d_A \leftarrow hash(b_A)$

（注17）：Mysten Labs, "GC depth," GitHub, https://github.com/YuseiWhite/sui/blob/mainnet/narwhal/Docker/validators/parameters.json#L3, February 2025.

（注18）：Mysten Labs, "TransactionValidator::validate function," GitHub, https://github.com/YuseiWhite/sui/blob/mainnet/narwhal/worker/src/tx_validator.rs#L16, February 2025.

（6） $w_{1,1} \in v_1$ はプライマリノードである $p \in v_1$ にダイジェスト $d_A$ を送信する．

（7） $payloadReceiver \in p_1$ は他のバリデータからもダイジェスト $D = \{d_B, d_C, \ldots, d_\omega\}$, $d_i \in \mathbb{N}$, $B \leq i \leq \omega$ を受信する．$proposer \in p$ は十分なダイジェスト $D$ (default: $D \geq 32$ (注19) (注20), $\max(D) = 1000$)(注21) (注22)があり，親の証明書 $cert_{Parents}$ が Quorum を達成している場合，Proposer::make_header() 関数(注23) (注24)により新たにヘッダー（認証前ブロック）$h_A$ (注25)を生成する．ヘッダー $h_A$ はワーカーのトランザクションバッチの概要や参照のための前ラウンド $r-1$ の $n-f$ の証明書を持っていて，信頼できるブロードキャスト [46] のための完全性と可用性を保証する．

（8） $Validator_1$ は $\{v \in V \mid v \neq v_1\}$ に対して $h_A$ をブロードキャストする．

（9） $\{v \in V \mid v \neq v_1\}, \exists r : r = Receiver(p(v))$ は受信した $h_A$ について以下を検証する．

- $h_A$ は正しいエポック [47] のものであるか
- $h_A$ のダイジェストが適切に作られたものであるか
- ワーカーノード内にメッセージ中で指定されたダイジェストに対応するバッチが保存されているか $h_A$ がラウンド $r$ でのジェネシスブロックであるかやラウンド $r-1$ から $2f+1$ 個の証明書が含まれているか等についてはヘッダー生成時に既に検証されている．

（10） $\{v \in V \mid v \neq v_1\}$ は，$h_A$ が有効である場合，プライマリノード $p_1 \in v$ は $h_A$ に署名することで投票する．VoteV1 構造体(注26)でヘッダーダイジェストに対する署名が必要であることが記されており，VoteV1::new() 関数(注27)によって署名と投票が実行される．

（11） $p_1 \in v_1$ が提案した $h_A$ が異なる受信者である他のバリデータから 2f+1 署名を集めることによって，クォーラムを達成し，証明書（認証済みブロック）$cert_A$ (注28)を生成する．この $cert_A$ は次のラウンド r+1 で生成されるであろう $cert_B$ の親ノードとして機能し，DAG 構築に不可欠である．また，証明書は，検証されたことを証明することを含むコンセンサスプロトコル内でのトランザクションまたはブロックの検証とファイナリティの達成にも不可欠である．

（12） $p_1 \in v_1$ は，$\{v \in V \mid v \neq v_1\}$, $\exists! p \in v$ に対してに $cert_A$ をブロードキャストする．これにより，受信者は $cert_A$ と関連するダイジェストを使用して，それぞれローカル DAG に頂点を作成できるようになる．

（13） プライマリノードが $n-f$ 個の頂点を形成し，それぞれが証明書を持つようになると，次のラウンド $r+1$ に進むことができる．このように形成された各ラウンドで各バリデータによる証明書 $cert_n$（コミット前ブロックまたは提案されたブロック）は Bullshark コンセンサスに渡されるサブ DAG として，最新ラウンドから降順でキュー構造でシーケンスされる．

**Consensus レイヤー: Total Ordering アルゴリズム.**

（14） $wave_{w-1}$ の $round_{max_r} \wedge r \in 2\mathbb{Z}$ で選出された Steady-State Leader と Fallback Leader が選出されている．

（15） Narwhal によってすでに処理されたサブ DAG を対象に最後にコミットされたサブダグ $cert_{committed}$ とリーダースケジュールをリカバーのために取得する，(注29) (注30) (注31) **?**

注意: ここから Bullshark コンセンサスプロトコルのプロセス開始が開始される．

（16） 新しい Bullshark 構造体(注32)のインスタンスを生成する．(注33)

（17） コンセンサスプロトコルでの処理を開始し(注34) (注35)，バリデータが新しい証明書を受信することで，Narwhal でのサブ

(注19)：Mysten Labs, "Batch size," GitHub, https://github.com/YuseiWhite/sui/blob/mainnet/narwhal/Docker/validators/parameters.json#L4, February 2025.

(注20)：Mysten Labs and Facebook, "Parameters::default_max_header_num_of_batches function," GitHub, https://github.com/YuseiWhite/sui/blob/mainnet/narwhal/config/src/lib.rs#L198-L200, February 2025.

(注21)：Mysten Labs, "Max header num of batches," GitHub, https://github.com/YuseiWhite/sui/blob/mainnet/narwhal/Docker/validators/parameters.json#L5, February 2025.

(注22)：Mysten Labs and Facebook, "Parameters::default_max_header_delay function," GitHub, https://github.com/YuseiWhite/sui/blob/mainnet/narwhal/config/src/lib.rs#L202-L204, February 2025.

(注23)：Mysten Labs and Facebook, "Proposer::make_header function," GitHub, https://github.com/YuseiWhite/sui/blob/mainnet/narwhal/primary/src/proposer.rs#L178-L200, February 2025.

(注24)：Mysten Labs and Facebook, "HeaderV1::new function," GitHub, https://github.com/YuseiWhite/sui/blob/mainnet/narwhal/types/src/primary.rs#L630-L649, February 2025.

(注25)：Mysten Labs and Facebook, "HeaderV1 struct," GitHub, https://github.com/YuseiWhite/sui/blob/mainnet/narwhal/types/src/primary.rs#L544-L556, February 2025.

(注26)：Mysten Labs and Facebook, "VoteV1 struct," GitHub, https://github.com/YuseiWhite/sui/blob/mainnet/narwhal/types/src/primary.rs#L961-L971, February 2025.

(注27)：Mysten Labs and Facebook, "VoteV1::new function," GitHub, https://github.com/YuseiWhite/sui/blob/mainnet/narwhal/types/src/primary.rs#L995-L1012, February 2025.

(注28)：Mysten Labs and Facebook, "CertificateV2 struct," GitHub, https://github.com/YuseiWhite/sui/blob/mainnet/narwhal/types/src/primary.rs#L1500-L1506, February 2025.

(注29)：Mysten Labs, "get_restored_consensus_output function," GitHub, https://github.com/YuseiWhite/sui/blob/mainnet/narwhal/executor/src/lib.rs#L90-L126, February 2025.

(注30)：Mysten Labs, "PrimaryNodeInner::start function," GitHub, https://github.com/YuseiWhite/sui/blob/mainnet/narwhal/node/src/primary_node.rs#L52-L113, February 2025.

(注31)：Mysten Labs, "PrimaryNodeInner::spawn_primary function," GitHub, https://github.com/YuseiWhite/sui/blob/mainnet/narwhal/node/src/primary_node.rs#L178-L283, February 2025.

(注32)：Mysten Labs and Facebook, "Bullshark struct," GitHub, https://github.com/YuseiWhite/sui/blob/mainnet/narwhal/primary/src/consensus/bullshark.rs#L26-L44, February 2025.

(注33)：Mysten Labs and Facebook, "Bullshark::new function," GitHub, https://github.com/YuseiWhite/sui/blob/mainnet/narwhal/primary/src/consensus/bullshark.rs#L48-L66, February 2025.

(注34)：Mysten Labs and Facebook, "Consensus struct," GitHub, https://github.com/YuseiWhite/sui/blob/mainnet/narwhal/primary/src/consensus/state.rs#L296-L320, February 2025.

(注35)：Mysten Labs and Facebook, "Consensus::run function," GitHub, https://github.com/YuseiWhite/sui/blob/mainnet/narwhal/primary/src/consensus/state.rs#L383-L391, February 2025.

DAG となる $cert_A$ が Bullshark に渡される．[注36][注37]

（18） Bullshark::process_certificate() 関数[注38]に対して ConsensusState 構造体[注39]を持つ Consensus 構造体[注40]とその証明書 $cert_A$ を渡す．

（19） $wave_w$ で $round_r = 1$ で $AnchorLeader_{A1}$ が配置され，全てのバリデータへブロードキャストされる．

（20） $round_{r+1} = 2$ で $n = 4$，$Validators = \{v_1, v_2, v_3, v_4\}$，$faulty_f = 1$ とする．$v1$ と $v2$ がそれぞれ $AnchorLeader_{A1}$ に投票する．

（21） $AnchorLeader_{A1}$ が Quorum（$f+1$）を達成するとこれが他の全てのバリデータに同期される．この結果から，$round_r$ の $AnchorLeader_{A1}$ は Steady-State Leader と定義される．

（22） $round_{r+2} = 3$ のための Steady-State Leader である $AnchorLeader_{A2}$ が選出される．

（23） $round_{r+2} = 3$ で $cert_A$ にすでに投票している $AnchorLeader_{A2}$ が配置され全てのバリデータへブロードキャストされる．

（24） $round_{r+3} = 4$ で $v_2$ と $v_3$ と $v_4$ が $AnchorLeader_{A2}$ に投票する．

（25） $(round_{r+3} = 4) \in 2\mathbb{Z}$ であることを確認し，最新のラウンド $round_{theLatest_r}$ と現在の ConsensusState 構造体[注41]を引数として Bullshark::commit_leader() 関数[注42]に渡す．[注43]

（26） $AnchorLeader_{A2}$ が Quorum（$f+1$）を達成し，これが他の全てのバリデータと同期される．

（27） $wave_{w+1}$ で投票対象となる Steady-State Leader と Fallback Leader が選出される．

（28） $round_{r+3} = 4$ の 1 つ前のラウンド（子ラウンド）$round_{r+2} = 3$ においての Quorum（$f+1$）を達成していることを検証する．[注44]

（29） Bullshark::commit_leader() 関数[注45]が Bullshark::process_certificate() 関数[注46]内で受け取った引数をそのまま Bullshark::order_leaders() 関数[注47]に渡し，最新のラウンド-2 から最後にコミットされたラウンド +2 の範囲 $x$（$round_{latest_r} - 2 \leq x \leq round_{lastCommitted_r} + 2$）を指定する．これは，ラウンドを 2 つずつ戻って最後にコミットされたラウンド +2 まで行うことで，例えば始めに $round_r$ と $round_{r-2}$ を対象にした場合の次は，$round_{r-2}$ と $round_{r-4}$ である．

（30） Bullshark::linked() 関数[注48]で 2 つの証明書（一番初めのループ処理では最新のラウンドのリーダーと一つ前のリーダーの間）が連結されているかどうかを確認する．ここでは，$AnchorLeader_{A1}$ と $AnchorLeader_{A2}$ が連結されているかを確認する．これらが連結されていれば，推移性があり，確率的な問題を完全に払拭できないものの，DAG ベースでのファイナリティ達成と見做すことができる可能性がある．

（31） アンカーリーダー（証明書）間が連結されている場合，コミット予定のアンカーリーダーを両端キューで Bullshark::commit_leader() 関数[注49]に返す．[注50] アンカーリーダーをコミットするということはその依存関係や因果履歴に対しても全てコミットするということを意味する．

（32） $AnchorLeader_{A1}$ と $AnchorLeader_{A2}$ は Quorum Intersection と DAG の構造により，$AnchorLeader_{A1}$ の後に $AnchorLeader_{A2}$ がコミットされるように順序付けされることが保証され，アンカーリーダーを古い順から順にシーケンスする．[注51]

（33） $cert_A$ を含むこれらのアンカーリーダーの順序付けありの集合をサブ DAG として他の全てのバリデータが保持できるように永続化する．[注52]

---

（注36）：Mysten Labs and Facebook, "Consensus::run_inner function," GitHub, https://github.com/YuseiWhite/sui/blob/mainnet/narwhal/primary/src/consensus/state.rs#L393-L476, February 2025.

（注37）：Mysten Labs and Facebook, "Consensus::spawn function," GitHub, https://github.com/YuseiWhite/sui/blob/mainnet/narwhal/primary/src/consensus/state.rs#L324-L381, February 2025.

（注38）：Mysten Labs and Facebook, "Bullshark::process_certificate function," GitHub, https://github.com/YuseiWhite/sui/blob/mainnet/narwhal/primary/src/consensus/bullshark.rs#L124-L214, February 2025.

（注39）：Mysten Labs and Facebook, "ConsensusState struct," GitHub, https://github.com/YuseiWhite/sui/blob/mainnet/narwhal/primary/src/consensus/state.rs#L36-L51, February 2025.

（注40）：Mysten Labs and Facebook, "Consensus struct," GitHub, https://github.com/YuseiWhite/sui/blob/mainnet/narwhal/primary/src/consensus/state.rs#L296-L320, February 2025.

（注41）：Mysten Labs and Facebook, "ConsensusState struct," GitHub, https://github.com/YuseiWhite/sui/blob/mainnet/narwhal/primary/src/consensus/state.rs#L36-L51, February 2025.

（注42）：Mysten Labs and Facebook, "Bullshark::commit_leader function," GitHub, https://github.com/YuseiWhite/sui/blob/mainnet/narwhal/primary/src/consensus/bullshark.rs#L221-L317, February 2025.

（注43）：Mysten Labs and Facebook, "Bullshark::process_certificate function," GitHub, https://github.com/YuseiWhite/sui/blob/mainnet/narwhal/primary/src/consensus/bullshark.rs#L124-L214, February 2025.

（注44）：Mysten Labs and Facebook, "Bullshark::commit_leader function," GitHub, https://github.com/YuseiWhite/sui/blob/mainnet/narwhal/primary/src/consensus/bullshark.rs#L221-L317, February 2025.

（注45）：Mysten Labs and Facebook, "Bullshark::commit_leader function," GitHub, https://github.com/YuseiWhite/sui/blob/mainnet/narwhal/primary/src/consensus/bullshark.rs#L221-L317, February 2025.

（注46）：Mysten Labs and Facebook, "Bullshark::process_certificate function," GitHub, https://github.com/YuseiWhite/sui/blob/mainnet/narwhal/primary/src/consensus/bullshark.rs#L124-L214, February 2025.

（注47）：Mysten Labs and Facebook, "Bullshark::order_leaders function," GitHub, https://github.com/YuseiWhite/sui/blob/mainnet/narwhal/primary/src/consensus/bullshark.rs#L321-L373, February 2025.

（注48）：Mysten Labs and Facebook, "Bullshark::linked function," GitHub, https://github.com/YuseiWhite/sui/blob/mainnet/narwhal/primary/src/consensus/bullshark.rs#L376-L392, February 2025.

（注49）：Mysten Labs and Facebook, "Bullshark::commit_leader function," GitHub, https://github.com/YuseiWhite/sui/blob/mainnet/narwhal/primary/src/consensus/bullshark.rs#L221-L317, February 2025.

（注50）：Mysten Labs and Facebook, "Bullshark::linked function," GitHub, https://github.com/YuseiWhite/sui/blob/mainnet/narwhal/primary/src/consensus/bullshark.rs#L376-L392, February 2025.

（注51）：Mysten Labs and Facebook, "Bullshark::commit_leader function," GitHub, https://github.com/YuseiWhite/sui/blob/mainnet/narwhal/primary/src/consensus/bullshark.rs#L221-L317, February 2025.

（注52）：Mysten Labs and Facebook, "Bullshark::commit_leader function," GitHub, https://github.com/YuseiWhite/sui/blob/mainnet/narwhal/primary/src/consensus/bullshark.rs#L221-L317, February 2025.

（34） コミットのトリガーをしたことの通知（Outcome::Commit）とトランザクション実行に必要な情報を保持しコンセンサスコミットするためのサブ DAG の順序付けありの集合(注53)を返す．(注54)(注55) この集合(注56)が決定論的ルールによってトランザクションの順序付けが行われる．

（35） トランザクションを実行するクライアントを生成する．(注57)(注58)(注59)

（36） サブ DAG 内の各証明書のすべてのバッチを保持する Consensus の出力(注60)を ConsensusHandler(注61)に渡す．(注62)(注63)

（37） Consensus の出力(注64)に対して以下についての全ての確認をとれた場合，トランザクションの順序付けを続ける．(注65)

- 最後にコミットされたラウンドと Consensus の出力内のリーダーラウンドと違うラウンドであること
- トランザクションの重複していないこと
- すでに処理されたトランザクションがないこと

（38） コミットされた全てのサブ DAG をもとにバッチ $b_A$ からトランザクション $tx_A$ が取り出され，実行トランザクションをガスフィーに従って決定論的に順序付け(注66)を行う．アンカー間の並び替えもトランザクションベースではここで行われる．(注67) トポロジカルソートによって複数の有効な順序が導出されることがあるが，異なるバリデータ間でも同じ順序でトランザクションを実行する合意とその一貫性を保証しているため，実行において問題になることはない．

（39） コミットに失敗した証明書は Narwhal に返す．(注68)

**Execution レイヤー: トランザクション実行アルゴリズム．**

（40） SuiNode が起動されており(注69)(注70)，プロトコルやメトリクス，TransactionManager(注71)等も初期化されている状態(注72)から，準備完了の証明書の実行処理の行うタスクを開始している(注73)．

（41） $tx_A$ などの新しいトランザクションが処理可能になるたびにループ処理を実行する，実行中に恒久的なエラーは起こることは基本的になく，一時的な障害のために最大 10 回，証明書を実行するロジック処理を行う AuthorityState::try_execute_immediately() 関数(注74)を実行，再試行する．再試行のインターバルは 1 秒とする．(注75)

ここでは，以下のことを保証している．

- コミット対象のオブジェクトはすでにロックされていること

オブジェクトのロックは enqueue_impl() 関数(注76)で実行されており，そのプロセスは以下の通りである．(注77)(注78)(注79)

（a） $tx_A$ を含むまだ実行されていないトランザクションだけを集める．

（b） 対象トランザクション $tx_A$ 内のオブジェクト情報を取

（注53）：Mysten Labs, "CommittedSubDag struct," GitHub, https://github.com/YuseiWhite/sui/blob/mainnet/narwhal/types/src/consensus.rs#L57-L72, February 2025.

（注54）：Mysten Labs and Facebook, "Bullshark::process_certificate function," GitHub, https://github.com/YuseiWhite/sui/blob/mainnet/narwhal/primary/src/consensus/bullshark.rs#L124-L214, February 2025.

（注55）：Mysten Labs and Facebook, "Bullshark::commit_leader function," GitHub, https://github.com/YuseiWhite/sui/blob/mainnet/narwhal/primary/src/consensus/bullshark.rs#L221-L317, February 2025.

（注56）：Mysten Labs, "CommittedSubDag struct," GitHub, https://github.com/YuseiWhite/sui/blob/mainnet/narwhal/types/src/consensus.rs#L57-L72, February 2025.

（注57）：Mysten Labs and Facebook, "Consensus::spawn function," GitHub, https://github.com/YuseiWhite/sui/blob/mainnet/narwhal/primary/src/consensus/state.rs#L324-L381, February 2025.

（注58）：Mysten Labs, "Executor::spawn function," GitHub, https://github.com/YuseiWhite/sui/blob/mainnet/narwhal/executor/src/lib.rs#L49-L87, February 2025.

（注59）：Mysten Labs, "spawn_subscriber function," GitHub, https://github.com/YuseiWhite/sui/blob/mainnet/narwhal/executor/src/subscriber.rs#L51-L99, February 2025.

（注60）：Mysten Labs, "ConsensusOutput struct," GitHub, https://github.com/YuseiWhite/sui/blob/mainnet/narwhal/types/src/consensus.rs#L24-L28, February 2025.

（注61）：Mysten Labs, "ConsensusHandler," GitHub, https://github.com/YuseiWhite/sui/blob/mainnet/crates/sui-core/src/consensus_handler.rs, February 2025.

（注62）：Mysten Labs, "ConsensusHandler::handle_consensus_output function," GitHub, https://github.com/YuseiWhite/sui/blob/mainnet/crates/sui-core/src/consensus_handler.rs#L205-L209, February 2025.

（注63）：Mysten Labs, "run_notify function," GitHub, https://github.com/YuseiWhite/sui/blob/mainnet/narwhal/executor/src/subscriber.rs#L101-L118, February 2025.

（注64）：Mysten Labs, "ConsensusOutput struct," GitHub, https://github.com/YuseiWhite/sui/blob/mainnet/narwhal/types/src/consensus.rs#L24-L28, February 2025.

（注65）：Mysten Labs, "ConsensusHandler::handle_consensus_output function," GitHub, https://github.com/YuseiWhite/sui/blob/mainnet/crates/sui-core/src/consensus_handler.rs#L205-L209, February 2025.

（注66）：Mysten Labs, "ConsensusTransactionOrdering enum," GitHub, https://github.com/YuseiWhite/sui/blob/mainnet/crates/sui-protocol-config/src/lib.rsL367-L373, February 2025.

（注67）：Mysten Labs, "ConsensusHandler::handle_consensus_output function," GitHub, https://github.com/YuseiWhite/sui/blob/mainnet/crates/sui-core/src/consensus_handler.rsL205-L209, February 2025.

（注68）：Mysten Labs and Facebook, "Consensus::run_inner function," GitHub, https://github.com/YuseiWhite/sui/blob/mainnet/narwhal/primary/src/consensus/state.rsL393-L476, February 2025.

（注69）：Mysten Labs, "main function in sui node," GitHub, https://github.com/YuseiWhite/sui/blob/mainnet/crates/sui-node/src/main.rs#L47-L163, February 2025.

（注70）：Mysten Labs, "SuiNode::start_async function," GitHub, https://github.com/YuseiWhite/sui/blob/mainnet/crates/sui-node/src/lib.rs#L390-L738, February 2025.

（注71）：Mysten Labs, "TransactionManager struct," GitHub, https://github.com/YuseiWhite/sui/blob/mainnet/crates/sui-core/src/transaction_manager.rs#L55-L63, February 2025.

（注72）：Mysten Labs and Facebook, "AuthorityState::new function," GitHub, https://github.com/YuseiWhite/sui/blob/mainnet/crates/sui-core/src/authority.rs#L2210-L2294, February 2025.

（注73）：Mysten Labs, "execution_process function," GitHub, https://github.com/YuseiWhite/sui/blob/mainnet/crates/sui-core/src/execution_driver.rs#L33-L123, February 2025.

（注74）：Mysten Labs and Facebook, "try_execute_immediately function," GitHub, https://github.com/YuseiWhite/sui/blob/mainnet/crates/sui-core/src/authority.rs#L952-L1001, February 2025.

（注75）：Mysten Labs, "execution_process function," GitHub, https://github.com/YuseiWhite/sui/blob/mainnet/crates/sui-core/src/execution_driver.rs#L33-L123, February 2025.

（注76）：Mysten Labs, "TransactionManager::enqueue_impl function," GitHub, https://github.com/YuseiWhite/sui/blob/mainnet/crates/sui-core/src/transaction_manager.rs#L394-L644, February 2025.

（注77）：Mysten Labs, "TransactionManager::enqueue function," GitHub, https://github.com/YuseiWhite/sui/blob/mainnet/crates/sui-core/src/transaction_manager.rs#L372-L379, February 2025.

（注78）：Mysten Labs, "TransactionManager::enqueue_certificates function," GitHub, https://github.com/YuseiWhite/sui/blob/mainnet/crates/sui-core/src/transaction_manager.rs#L359-L369, February 2025.

（注79）：Mysten Labs and Facebook, "AuthorityState::enqueue_certificates_for_execution function," GitHub, https://github.com/YuseiWhite/sui/blob/mainnet/crates/sui-core/src/authority.rs#L2322-L2329, February 2025.

得する．

（c） 対象オブジェクト $object_A$ の書き込みロックを取得する．

（d） 対象オブジェクト $object_A$ の可用性を確認する．

（e） 対象オブジェクト $object_A$ の可用性とキーをペアする．

（f） $cert_A$ などの実行可能な証明書を準備完了とし，ExecutionDriver に送信する．(注80)

ロックの設定の保証ができない場合は，execute_certificate() 関数(注81)を実行する

- 実行と出力のコミットは原子的（Atomic）に行われ，クラッシュした実行はどんな影響も及ぼさないこと

（42） $tx_A$ で実行したいトランザクションの効果がすでに書き込まれているかを確認し，同じトランザクションの重複実行を避けるようにする．(注82)

（43） トランザクション $tx_A$ の実行に必要な入力オブジェクト $object_A$ を読み取る．(注83)

（44） 実際の証明書の処理を開始し，トランザクション $tx_A$ の実行ロックを取得し，実行準備のための検証を行う．(注84)(注85)(注86) この検証は所有オブジェクトと共有オブジェクトの両方のロックについて，データベースから状態を読み取っているだけであるため，副作用はない．

（45） 対象トランザクション $tx_A$ をコミットする．(注87) これにより決定論的ファイナリティが達成される．

（46） $tx_A$ のトランザクションキーとその効果の署名をエポックストアに挿入する．(注88) これにより，トランザクション $tx_A$ の記録が永続化されるようになる．

（47） トランザクション $tx_A$ とその出力オブジェクト $object_{A'}$ がコミットされたことを TransactionManager(注89)に通知する．(注90)(注91)

（48） ロックしていたオブジェクト $object_A$ を解放する．(注92)

## 7. 結　論

Bullshark は，Narwhal 上に構築された，ゼロオーバーヘッドの BFT アプローチのビザンチンアトミックブロードキャストプロトコルであり，部分同期バージョンの最適化によって，ガベージコレクションと公平性の競合を解決した初めてのプロトコルである．また，Bullshark は DAG-Rider での部分同期中の Liveness やポスト量子セキュリティ，複雑性の排除などの特性を維持している．パフォーマンスにおいては，バリデータの増加に伴って線形的に TPS も増加し，50 台のバリデータ数では，約 130,000 tx/s と HotStuff の 2 倍のスループットを達成している．ビザンチン障害時においても約 100,000 tx/s，4 秒程度のレイテンシという高パフォーマンスを記録している．しかし，Sui Blockchain で採用されているプロダクション環境での Bullshark の平均レイテンシは 2.5 秒かかっていること，ビザンチン障害の多い場合（10 台のうち 3 台が faulty なバリデータ），60,000 tx/s ほどしか維持できないことについては，今後のパフォーマンス改善が期待される．この障害に対して，回復のためのオーバーヘッドが必要となっていることも改善の余地がある．アンカーリーダーのコミット提案について，$n-f$ 未満の投票の場合はコミットされないことから，このバリデータの貢献を活用できるようにすることが今後の研究で期待される．ガベージコレクションも一つの有効な解決策であるが，貢献が活用されるかは他のバリデータの投票次第となってしまっている．公平性も同期中は保証されないため，公平性の幅を拡大するようなシステム設計も期待されるだろう．公平性の観点においては分散性も重要となってくる．DAG はそのパフォーマンスの優位性から分散化について不完全であり [48]，Bullshark はワーカーノードの増加によってこの問題を緩和しているが，プライマリノードについては，同様の課題を抱えている．今後，現在のパフォーマンスを維持しながらも分散性も高める DAG ベースコンセンサスプロトコルの提案が期待される．また，ビザンチン障害下における BFT プロトコルの評価方法については未だに未解決の研究課題となっている．[49] 本研究では，プロダクション環境において実行アルゴリズムも含めた詳細な分析を提供しているが，今後プロダクション環境での研究の幅を広げ，ネッ

（注80）：Mysten Labs, "TransactionManager::certificate_ready function," GitHub, https://github.com/YuseiWhite/sui/blob/mainnet/crates/sui-core/src/transaction_manager.rs#L730-L742, February 2025.

（注81）：Mysten Labs and Facebook, "AuthorityState::execute_certificate function," GitHub, https://github.com/YuseiWhite/sui/blob/mainnet/crates/sui-core/src/authority.rs#L911-L938, February 2025.

（注82）：Mysten Labs and Facebook, "try_execute_immediately function," GitHub, https://github.com/YuseiWhite/sui/blob/mainnet/crates/sui-core/src/authority.rs#L952-L1001, February 2025.

（注83）：Mysten Labs and Facebook, "try_execute_immediately function," GitHub, https://github.com/YuseiWhite/sui/blob/mainnet/crates/sui-core/src/authority.rs#L952-L1001, February 2025.

（注84）：Mysten Labs and Facebook, "Bullshark::process_certificate function," GitHub, https://github.com/YuseiWhite/sui/blob/mainnet/narwhal/primary/src/consensus/bullshark.rs#L124-L214, February 2025.

（注85）：Mysten Labs and Facebook, "try_execute_immediately function," GitHub, https://github.com/YuseiWhite/sui/blob/mainnet/crates/sui-core/src/authority.rs#L952-L1001, February 2025.

（注86）：Mysten Labs and Facebook, "AuthorityState::prepare_certificate function," GitHub, https://github.com/YuseiWhite/sui/blob/mainnet/crates/sui-core/src/authority.rs#L1341-L1405, February 2025.

（注87）：Mysten Labs and Facebook, "AuthorityState::commit_certificate function," GitHub, https://github.com/YuseiWhite/sui/blob/mainnet/crates/sui-core/src/authority.rs#L1215-L1289, February 2025.

（注88）：Mysten Labs and Facebook, "AuthorityState::commit_certificate function," GitHub, https://github.com/YuseiWhite/sui/blob/mainnet/crates/sui-core/src/authority.rs#L1215-L1289, February 2025.

（注89）：Mysten Labs, "TransactionManager struct," GitHub, https://github.com/YuseiWhite/sui/blob/mainnet/crates/sui-core/src/transaction_manager.rs#L55-L63, February 2025.

（注90）：Mysten Labs and Facebook, "AuthorityState::commit_certificate function," GitHub, https://github.com/YuseiWhite/sui/blob/mainnet/crates/sui-core/src/authority.rs#L1215-L1289, February 2025.

（注91）：Mysten Labs, "TransactionManager::notify_commit function," GitHub, https://github.com/YuseiWhite/sui/blob/mainnet/crates/sui-core/src/transaction_manager.rs#L697-L727, February 2025.

（注92）：Mysten Labs, "TransactionManager::notify_commit function," GitHub, https://github.com/YuseiWhite/sui/blob/mainnet/crates/sui-core/src/transaction_manager.rs#L697-L727, February 2025.

トワーク環境の分析やセキュリティ分析を行うことも期待される．最後に，Bullshark は CAP 定理 [50] において，Consistency と Partial Tolerance を満たす因果履歴に基づく一貫性を保証しているが，Sui の Object-Centric Data Model [51] と組み合わせて高い Availability を実現している．ビザンチン環境下を除けば，Narwhal と Bullshark の組み合わせは，CAP 定理に対してブロックチェーン分野の実用面での最適解と考えられるかもしれない．これは今後のパフォーマンス向上によって解決策が更新されていくだろう．

## 文　献